\documentclass[3p,twocolumn]{elsarticle}
\pdfoutput=1
\DeclareGraphicsExtensions{.pdf}

\biboptions{sort&compress}
\usepackage{multirow}
\usepackage{amssymb}
\usepackage{amsmath}
\usepackage{slashed}
\usepackage{graphicx}
\usepackage{atlasphysics}
\usepackage{placeins}
\usepackage{array}
\newcommand{\papertype}{Letter}

\usepackage{preprintcover}  
\PreprintCoverPaperTitle{
Measurement of inclusive jet charged-particle fragmentation functions
in Pb+Pb collisions at $\sqrt{s_{NN}} = 2.76$~TeV with the ATLAS detector
}  

\PreprintIdNumber{CERN-PH-EP-2014-101} 

\PreprintCoverAbstract{
  Measurements of charged-particle fragmentation functions of jets produced in ultra-relativistic nuclear 
collisions can provide insight into the modification of parton showers in the hot, dense medium created in 
the collisions. ATLAS has measured jets in $\sqrt{s_{NN}} = 2.76$~TeV Pb+Pb collisions at the LHC using a 
data set recorded in 2011 with an integrated luminosity of $0.14~{\rm nb^{-1}}$. Jets were reconstructed 
using the \mbox{anti-$k_{t}$} algorithm with distance parameter values $R = 0.2, 0.3,$ and 0.4. 
Distributions of charged-particle transverse momentum and longitudinal momentum fraction are reported for 
seven bins in collision centrality for $R=0.4$ jets with $p_{{\mathrm{T}}}^{\mathrm{jet}}
> 100$~GeV. Commensurate minimum $p_{\mathrm{T}}$ values are used for the other radii.
  Ratios of fragment distributions in each centrality bin to those measured in the most peripheral bin are 
presented. These ratios show a reduction of fragment yield in central collisions relative to peripheral 
collisions at intermediate $z$ values, $0.04 \lesssim z \lesssim 0.2$ and an enhancement in fragment yield 
for $z \lesssim 0.04$. A smaller, less significant enhancement is observed at large $z$ and large 
$p_{\mathrm{T}}$ in central collisions.
  }  
\PreprintJournalName{Physics Letters B}

\newcommand{\PbPb}{\mbox{Pb+Pb}}
\newcommand{\pp}{\mbox{$pp$}}

\newcommand{\Ncone}{\mbox{$N_{\mathrm{cone}}$}}

\newcommand{\Njet}{\mbox{$N_{\mathrm{jet}}$}}

\newcommand{\Nch}{\mbox{$N_{\mathrm{ch}}$}}

\newcommand{\nchUE}{\mbox{$n_{\mathrm{ch}}^{\mathrm{UE}}$}}

\newcommand{\sqrtsnn}{\mbox{$\sqrt{s_{\mathrm{NN}}}$}}

\newcommand{\RFour}{\mbox{$R = 0.4$}}
\newcommand{\RThree}{\mbox{$R = 0.3$}}
\newcommand{\RTwo}{\mbox{$R= 0.2$}}

\newcommand{\antikt}{\mbox{anti-\kt}}
\newcommand{\avgpttrue}{\mbox{$\langle {p_{\mathrm{T}}}^{\mathrm{jet}}_{\mathrm{true}}\rangle$}}

\newcommand{\ETfcal}{\mbox{$\Sigma E_{\mathrm{T}}^{\mathrm{FCal}}$}}

\newcommand{\gjet}{\mbox{$\gamma$+jet}}
\newcommand{\invnb}{\mbox{${\rm nb^{-1}}$}}
\newcommand{\kt}{\mbox{$k_{t}$}}

\newcommand{\diff}{\mathrm{d}}

\newcommand{\Dz}{\mbox{$D(z)$}}
\newcommand{\dz}{\mathrm{d}z}
\newcommand{\Dzmeas}{\mbox{$D^{\mathrm{meas}}(z)$}}

\newcommand{\Dpt}{\mbox{$D(\pT)$}}
\newcommand{\Dptmeas}{\mbox{$D^{\mathrm{meas}}(\pT)$}}

\newcommand{\Dptratio}{\mbox{$D(\pT)|_{\mathrm{cent}}/D(\pT)|_{\mathrm{60-80}}$}}
\newcommand{\Dzratio}{\mbox{$D(z)|_{\mathrm{cent}}/D(z)|_{\mathrm{60-80}}$}}

\newcommand{\effpteta}{\mbox{$\varepsilon(\pT, \eta)$}}

\newcommand{\pthat}{\mbox{$\hat{p}_{\mathrm{T}}$}}
\newcommand{\etajet}{\mbox{$\eta^{\mathrm{jet}}$}}
\newcommand{\phijet}{\mbox{$\phi^{\mathrm{jet}}$}}
\newcommand{\pTjet}{\mbox{$p_{{\mathrm{T}}}^{\mathrm{jet}}$}}

\newcommand{\pTch}{\mbox{$p_{{\mathrm{T}}}^{\mathrm{ch}}$}}

\newcommand{\pTrec}{\mbox{${p_{{\mathrm{T}}}}^{\mathrm{jet}}_{\mathrm{rec}}$}}
\newcommand{\pTtrue}{\mbox{${p_{\mathrm{T}}}^{\mathrm{jet}}_{\mathrm{true}}$}}
\newcommand{\pTchtrue}{\mbox{${p_{\mathrm{T}}}^{\mathrm{ch}}_{\mathrm{true}}$}}
\newcommand{\pTchrec}{\mbox{${p_{\mathrm{T}}}^{\mathrm{ch}}_{\mathrm{rec}}$}}
\newcommand{\ztrue}{\mbox{$z^{\mathrm{true}}$}}
\newcommand{\zrec}{\mbox{$z^{\mathrm{rec}}$}}
\newcommand{\ptvjet}{\mbox{$\displaystyle {\vec{p}_{\mathrm{T}}}^{\, \mathrm{jet}}$}}
\newcommand{\ptvch} {\mbox{$\displaystyle {\vec{p}_{\mathrm{T}}}^{\, \mathrm{ch}}$}}

\newcommand{\Rdz}{\mbox{$R_{D(z)}$}}
\newcommand{\Rdpt}{\mbox{$R_{D(\pT)}$}}
\newcommand{\Rdzmeas}{\mbox{$R_{D(z)}^{\mathrm{meas}}$}}
\newcommand{\Rdptmeas}{\mbox{$R_{D(\pT)}^{\mathrm{meas}}$}}

\newcommand{\ARdz}{\overline{R}_{D(z)}}
\newcommand{\ARdpt}{\overline{R}_{D(\pT)}}
\newcommand{\SARdz}{\sigma(\overline{R}_{D(z)})}
\newcommand{\SARdpt}{\sigma(\overline{R}_{D(\pT)})}

\begin{document}

\title{
\center{Measurement of inclusive jet charged particle fragmentation functions
  in \PbPb\ collisions at $\sqrtsnn = 2.76$~\TeV\ with the ATLAS detector}
}

\author{The ATLAS Collaboration}

\begin{abstract}
Measurements of charged-particle fragmentation functions of jets produced in 
ultra-relativistic nuclear collisions can provide insight into the 
modification of parton showers in the hot, dense medium created in the collisions. 
ATLAS has measured jets in $\sqrtsnn = 2.76$~\TeV\ \PbPb\ 
collisions at the LHC using a data set recorded in 2011 with an integrated luminosity of 
$0.14~\invnb$. Jets were reconstructed using the \antikt\ algorithm with distance 
parameter values $R = 0.2, 0.3,$ and 0.4. Distributions of charged-particle 
transverse momentum 
 and longitudinal momentum fraction 
 are reported for seven bins in collision centrality for 
\RFour\ jets with $\pTjet > 100$~GeV. Commensurate minimum \pT\ values
are used for the other radii. 
Ratios of fragment distributions in each centrality bin
to those measured in the most peripheral bin are presented. 
These ratios show a reduction of fragment yield in central collisions 
relative to peripheral collisions at intermediate $z$ values, $0.04 \lesssim z \lesssim 
0.2$ and an enhancement in fragment yield for $z \lesssim 0.04$. A
smaller, less significant enhancement is observed at large $z$ and
large \pt\ in central collisions.

\end{abstract}
\maketitle 

\newlength{\fighalfwidth}
\setlength{\fighalfwidth}{0.49\textwidth}

\section{Introduction}
Collisions between lead nuclei at the LHC are thought to produce
a quark--gluon plasma (QGP), a form of strongly interacting matter in
which quarks and gluons become locally deconfined. One predicted
consequence of QGP formation is the ``quenching'' of jets generated
in hard-scattering processes during the initial stages of the nuclear 
collisions \cite{Majumder:2010qh}. Jet quenching refers, collectively,
to a set of possible modifications of parton showers by the QGP
through interactions of the constituents of the shower with the colour
charges in the plasma \cite{Mehtar-Tani:2013pia,Armesto:2011ht}. 
In particular, quarks and
gluons in the shower may be elastically or inelastically scattered
resulting in both deflection and energy loss of the constituents of
the shower. The deflection and the extra radiation associated with
inelastic processes may broaden the parton shower and eject partons out
of an experimental jet cone
\cite{Qin:2010mn,Ovanesyan:2011kn,CasalderreySolana:2011rq,CasalderreySolana:2012ef,Blaizot:2013hx,Casalderrey-Solana:2014bpa}. As a result, jet quenching
can  potentially both soften the spectrum of the momentum of hadrons inside the jet and reduce
the total energy of the reconstructed jet. A complete characterization
of the effects of jet quenching therefore requires measurements of both 
the single-jet suppression and the jet fragment
distributions. 

Observations of modified dijet asymmetry distributions 
\cite{Aad:2010bu,Chatrchyan:1327643,Chatrchyan:1426312}, modified balance-jet 
transverse momentum (\pT) distributions in \gjet\ events \cite{Chatrchyan:2012gt}, 
and suppressed inclusive jet yield in \PbPb\ collisions at the LHC \cite{jetsupp,Abelev:2013kqa} 
are consistent with theoretical calculations of jet quenching. However, it has been 
argued that those measurements do not sufficiently discriminate between 
calculations that make different assumptions regarding the relative importance of 
the contributions described above \cite{Renk:2012cx}. 
  Based on the above arguments, theoretical analyses are incomplete without 
experimental constraints on the theoretical description of jet fragment 
distributions.

  This \papertype\ presents measurements of charged-particle jet
fragmentation functions in  $\sqrtsnn = 2.76$~\TeV\ \PbPb\ collisions
using $0.14~\invnb$ of data  recorded in 2011. The jets used in the 
measurements were reconstructed with the
\antikt\ \cite{Cacciari:2008gp} algorithm using 
distance parameter values $R = 0.2, 0.3,$ and 0.4. 
    Results are
  presented for the charged-particle transverse momentum (\ptvch) and
  longitudinal momentum  fraction ($z \equiv \ptvch \cdot
  \ptvjet/|\ptvjet|^2$) distributions,  
  \begin{equation}
\Dpt \equiv \dfrac{1}{\Njet} \dfrac{\diff\Nch}{\diff\pTch},\\
  \end{equation}
  \begin{equation}
\Dz \equiv \dfrac{1}{\Njet} \dfrac{\diff\Nch}{\diff z},
\label{eq:ffuncdef}
  \end{equation}
   of charged particles with $\pTch > 2$~\GeV\ produced within an angular
range $\Delta R = 0.4$ 
of the reconstructed jet directions for jets with $\pTjet > 85, 92,$
   and 100~\GeV\ for $R = 0.2, 0.3,$ and 0.4, respectively. Here, $\Delta
R = \sqrt{(\Delta \phi)^2 + (\Delta \eta)^2}$ where $\Delta \phi$
($\Delta  \eta$) is the difference in azimuthal angles
(pseudorapidities) between the charged particle and 
jet directions.\footnote{
  ATLAS uses a  
right-handed coordinate system with its origin at the nominal
interaction point (IP) in the centre of the detector and the
$z$-axis along the beam pipe. The $x$-axis points from the IP to the
centre of the LHC ring, and the $y$ axis points upward. Cylindrical
coordinates $(r,\phi)$ are used in the transverse plane, $\phi$
being the azimuthal angle around the beam pipe. The pseudorapidity
is defined in terms of the polar angle $\theta$ as
$\eta=-\ln\tan(\theta/2)$.}
  The \pTjet\ thresholds for the three $R$ values were chosen to match the $R$-dependence of the measured 
transverse momentum of a typical jet. For simplicity, the terms ``fragmentation functions'' are used to 
describe the distributions defined in Eq.~(\ref{eq:ffuncdef}) with the understanding that \Dz\ is different 
from a theoretical fragmentation function, $D(z,Q^2)$, 
calculated using unquenched jet energies and with no restriction on the angles of particles with respect to the jet axis.
  Earlier measurements by CMS of jet fragmentation functions
\cite{Chatrchyan:2012gw} in \PbPb\ collisions at the LHC show no significant 
modification, but the uncertainties on that measurement were not sufficient to 
exclude modifications at the level of $\sim 10\%$. 
  CMS recently released a new result {\mbox{\cite{Chatrchyan:2014ava}}} using higher statistics data from
2011 that show fragmentation function modifications which are consistent with the results presented in this
Letter.

\section{Experimental setup}
\label{sec:experiment}
The measurements presented in this \papertype\ were performed using the ATLAS
calorimeter, inner detector, muon spectrometer, trigger, and data acquisition systems
\cite{Aad:2008zzm}. The ATLAS calorimeter system consists of a liquid
argon (LAr) electromagnetic (EM) calorimeter covering $|\eta|<3.2$, a
steel--scintillator sampling hadronic calorimeter covering $|\eta| <
1.7$, a LAr hadronic calorimeter covering $1.5 < |\eta| < 3.2$, and
two LAr forward calorimeters (FCal) covering $3.2 < |\eta| <
4.9$. The hadronic calorimeter has three
sampling layers longitudinal in shower depth and has a $\Delta
\eta \times \Delta \phi$ granularity of $0.1 \times
0.1$ for $|\eta| < 2.5$ and  $0.2 \times 0.2$ for $2.5 < |\eta| < 4.9$.\footnote{An 
  exception is the third sampling layer that has a segmentation of $0.2 \times 0.1$
up to $|\eta| = 1.4$.}
The EM calorimeters are segmented longitudinally in shower depth into three
compartments with an additional pre-sampler layer. The EM calorimeter 
has a granularity that varies with layer and pseudorapidity, but which
is generally much finer than that of the hadronic
calorimeter. The middle sampling layer, which typically has the
largest energy deposit in EM showers, has a granularity of $0.025
\times 0.025$ over  $|\eta| < 2.5$. 

The inner detector \cite{Collaboration:2010knc} measures charged particles 
within the pseudorapidity interval 
$|\eta|<2.5$ using a combination of silicon pixel detectors, silicon
microstrip detectors (SCT), and a straw-tube transition radiation
tracker (TRT),
all immersed in a 2~T axial magnetic field. All three detectors are
composed of a barrel and two symmetrically placed end-cap sections.
The pixel detector is composed of 3 layers of sensors with nominal
feature size $50~{\rm \mu m} \, \times \, 400~{\rm \mu m}$.  The 
SCT barrel section contains 4 layers of modules with 80~$\mu {\rm
  m}$ pitch sensors on both sides, while each end-cap consists of nine
layers of double-sided modules with radial strips having a mean pitch
of $80~\mathrm{\mu m}$. The two sides of each SCT layer in both the barrel
and the end-caps have a relative stereo angle of 40~mrad. The
TRT contains up to 73 (160) layers of
staggered straws interleaved with fibres in the barrel (end-cap).
Charged particles with $\pTch \gtrsim 0.5$~\GeV\ 
typically traverse three layers of pixel sensors, four
layers of double-sided SCT sensors, and, in the case of $|\eta|<2.0$, 36 TRT straws.

Minimum bias \PbPb\ collisions were identified using measurements from
the zero degree calorimeters (ZDCs) and the minimum-bias trigger
scintillator (MBTS) counters \cite{Aad:2008zzm}. The ZDCs are located
symmetrically at $z = \pm 140$~m and cover $|\eta| > 8.3$. In
\PbPb\ collisions the ZDCs measure primarily ``spectator'' neutrons,
which originate from the incident nuclei and do not interact
hadronically. 
   The MBTS detects charged particles over $2.1 < |\eta| < 3.9$ using two 
 counters placed at $z = \pm 3.6$~m.  
  MBTS counters are divided into 16 
modules with 8 different positions in azimuth and covering 2 different 
$|\eta|$ intervals. Each counter provides measurement of both the pulse 
heights and arrival times of ionization energy deposits.

  Events used in this analysis were selected for recording by a
combination of Level-1 minimum-bias and High Level Trigger (HLT) jet
triggers. 
  The Level-1 trigger required 
  a total transverse energy measured in the calorimeter of greater than 10~\GeV.
The HLT jet trigger ran the offline \PbPb\ jet reconstruction algorithm,
described below, for \RTwo\ jets except for the application of the
final hadronic energy scale correction. The HLT trigger selected events
containing an \RTwo\ jet with transverse energy $\ET >
20$~\GeV.     

\section{Event selection and data sets}
\label{sec:analysis}
  This analysis uses a total integrated luminosity of $0.14~\invnb$ of \PbPb\ 
collisions recorded by ATLAS in 2011. Events selected by the HLT jet
trigger were required to have a
reconstructed primary vertex and a  time difference between hits in
the two sides of the MBTS detector of less than 3~ns.  
The primary vertices were reconstructed from charged-particle tracks with 
\mbox{$\pTch > 0.5$~\GeV}. The tracks were reconstructed from hits in the inner 
   detector using the ATLAS track reconstruction algorithm described in Ref.~\cite{Aad:2010rd} 
   with settings optimized for the high hit density in 
heavy-ion collisions \cite{ATLAS:2011ah}. A total of 14.2 million events
passed the described selections.

The centrality of \PbPb\ collisions was characterized by \ETfcal, the total 
transverse energy measured in the forward calorimeters \cite{ATLAS:2011ah}. 
Jet fragmentation functions were measured in seven centrality bins defined 
according to successive percentiles of the \ETfcal\ distribution ordered from 
the most central to the most peripheral collisions: \mbox{0--10\%}, 
\mbox{10--20\%}, \mbox{20--30\%}, \mbox{30--40\%}, \mbox{40--50\%}, 
\mbox{50--60\%}, and \mbox{60--80\%}. The percentiles were defined after 
correcting the \ETfcal\ distribution for a 2\% minimum-bias trigger 
inefficiency that affects the most peripheral events which are not included 
in this analysis.

The performance of the ATLAS detector and offline analysis in
measuring jets and charged particles in the environment of
\PbPb\ collisions was evaluated using a large Monte Carlo
(MC) event sample obtained by overlaying simulated \cite{atlassim} 
PYTHIA \cite{Sjostrand:2006za} \pp\ hard-scattering events at $\sqrt{s}=2.76$~TeV onto 1.2 million
minimum-bias \PbPb\ events recorded in 2011. The same number of PYTHIA
events was produced for each of five intervals of \pthat, the
transverse momentum of outgoing partons in the $2\rightarrow 2$
hard-scattering, with boundaries $17, 35, 70, 140, 280,$ and
560~\GeV. The detector response to the PYTHIA events was simulated
using Geant4 \cite{Agostinelli:2002hh}, and the simulated hits were 
combined with the data from the minimum-bias \PbPb\ events to produce
1.2 million overlaid events for each \pthat\ interval.

\section{Jet and charged-particle analysis}
Charged particles included in the fragmentation measurements were
required to have at least two hits in the pixel detector, including a
hit in the first pixel layer
  if the track trajectory makes such a hit expected,
and seven hits in the silicon microstrip detector. In addition, the 
transverse ($d_0$) and longitudinal ($z_0\sin{\theta}$) 
impact parameters of the tracks measured with
respect to the primary vertex were required to satisfy
$|d_0/\sigma_{d_0} | < 3$ and 
$|z_0\sin{\theta}/\sigma_{z}| <3$, where $\sigma_{d_0}$ and $\sigma_{z}$ are  
uncertainties on $d_0$ and $z_0\sin{\theta}$, respectively, obtained
from the track-fit covariance matrix. 
 Jets were reconstructed using the techniques described in Ref.~\cite{jetsupp}, which are briefly summarized here.

The \antikt\ algorithm was first run in four-momentum recombination mode, on
$\Delta \eta \times \Delta \phi = 0.1\times 0.1$ logical towers and for
three values of the \antikt\ distance parameter, $R = 0.2, 0.3,$ and
0.4. The tower kinematics were obtained by summing electromagnetic-scale 
energies of calorimeter cells within the tower boundaries. Then,
  an iterative procedure was used to estimate a layer- and $\eta$-dependent 
underlying event (UE) energy density while excluding actual jets from that 
estimate. 
  The UE energy was subtracted from each calorimeter cell within the towers 
included in the reconstructed jet. 
  The correction takes into
acount a $\cos{2\phi}$ modulation of the calorimeter response due to 
elliptic flow of the medium \cite{ATLAS:2011ah} which is estimated by 
measurement of the amplitude of that modulation in the calorimeter.
  The final jet kinematics were calculated via a 
four-momentum sum of all (assumed massless) cells contained within the jets using 
subtracted \ET\ values.
  A correction was applied to the reconstructed jet to account
for jets not excluded or only partially excluded from the UE estimate.
  Then, a final jet $\eta$- and \ET-dependent hadronic energy 
scale calibration factor was applied.

 After the reconstruction, 
additional selections were applied for the
purposes of this analysis. ``UE jets'' generated by
fluctuations in the underlying event, were removed using techniques
described in Ref.~\cite{jetsupp}. 

  To prevent neighbouring jets from distorting the measurement of the
fragmentation functions, jets were required to be isolated. The
isolation cut required that there be no other jet within $\Delta R =
1$ having $\pt > p_{\mathrm{T}}^{\mathrm{iso}}$ where
$p_{\mathrm{T}}^{\mathrm{iso}}$, the isolation threshold, is set to 
half of the analysis threshold for each  $R$ value, 
$p_{\mathrm{T}}^{\mathrm{iso}} = 42.5, 46,$ and 50~\GeV\ 
for $R = 0.2, 0.3,$ and 0.4, respectively. To prevent muons 
from semileptonic heavy-flavour decays from influencing the measured
fragmentation functions, all jets with reconstructed muons having $\pt
> 4$~\GeV\ within a cone of size $\Delta R = 0.4$ were excluded from the
analysis. To prevent inactive regions in the calorimeters from
producing artificial high $z$ fragments, jets were required to have more than 90\%
of their energy contained within fully functional regions of the
calorimeter. Finally, all jets included in the analysis were required
to match HLT jets reconstructed 
  with transverse
momenta greater than the trigger threshold of 20~GeV. 
The HLT jets were found to be fully efficient for the jet kinematic 
selection used in this analysis.
  Table~\ref{tbl:cutflow}
  shows the impact of the cuts on the number of measured jets 
in central (\mbox{0--10\%}) and peripheral (\mbox{60--80\%}) collisions. 
All these cuts together retain more than 96\% of all jets.

 \begin{table}[t]
 \centering
 \begin{tabular}{| l ||  c | c | } \hline
                & \multicolumn{2}{c|}{\Njet} \\ \hline 
 Cut description & \mbox{0--10\%} &  \mbox{60--80\%}
 \\ \hline \hline
 All jets                  &  41191    &   2579   \\ \hline
 UE jet rejection          &  41116    &   2570   \\ \hline 
 Isolation                 &  40986    &   2554   \\ \hline
 Muon rejection            &  40525    &   2523   \\ \hline
 Inactive area exclusion   &  39548    &   2458   \\ \hline
 Trigger jet match         &  39548    &   2458   \\ \hline
 \end{tabular}
 \caption{
   Number of jets for two centrality bins in data as a function of the 
   selection criteria applied.
   Each line specifies 
   the number of jets passing all cuts for the given line and above. 
 }
 \label{tbl:cutflow}
 \end{table}

\section{Jet and track reconstruction performance}
\label{sec:perform}
The performance of the ATLAS detector and analysis procedures in
measuring jets was evaluated from the MC sample using
the procedures described in Ref.~\cite{jetsupp}. Reconstructed MC jets 
were matched to ``truth'' jets obtained
by separately running the \antikt\ algorithm on the final-state PYTHIA
particles\footnote{
  Final-state PYTHIA particles are defined as 
all generated particles with lifetimes longer than $0.3 \cdot
10^{-10}$~s originating from the primary interaction or from
subsequent decay of particles with shorter lifetimes.  
}
  for the three jet $R$ values used in this analysis. 
For the jet fragmentation measurements, the most important
aspect of the jet performance is the jet energy resolution (JER). For
jet energies $\gtrsim 100$~\GeV, the JER in central (\mbox{0--10\%})
collisions for \RFour\ jets has comparable contributions from UE
fluctuations  and ``intrinsic'' resolution of the calorimetric jet
measurement. For peripheral collisions and \RTwo\ jets, the intrinsic
calorimeter resolution dominates the JER. 
  The value of JER evaluated for jets with $\pt=100$~GeV in 0--10\% collisions is 0.18, 0.15, and 0.13 for $R=0.4$, $R=0.3$, and $R=0.2$ jets, respectively.

The combination of the finite JER and
the steeply falling jet \pt\ spectrum produces a net migration of jets from
lower \pt\ to higher \pt\ values (hereafter referred to as
``upfeeding'') such that a jet reconstructed with a given
\pTrec\ corresponds, on average, to a lower truth-jet
\pT, \avgpttrue. The relationship between \avgpttrue\ and \pTrec\ was
evaluated from the MC data set for the different centrality bins
and three $R$ values used in this analysis. For the jet
\pTrec\ values used in this analysis, that relationship is well
described by a linear dependence,  
\begin{equation}
\avgpttrue = \alpha \pTrec + \beta.
\label{eq:avgpttruedef}
\end{equation}
Sample values for $\alpha$ and $\beta$ and the resulting 
\avgpttrue\ values for \RTwo\ and \RFour\ jets in peripheral and
central collisions are listed in Table~\ref{tbl:avgpttrue}.
The extracted relationships between \pTrec\ and \avgpttrue\ will be
used in the fragmentation analysis to correct for the average
shift in the measured jet energy.
\begin{table*}
\centerline{
\begin{tabular}{|l|l||c|c|r|}\hline
Centrality & Jet $R$ & $\alpha$ & $\beta$ (GeV) & $\avgpttrue(100 \GeV)$
\\ \hline \hline
\mbox{0--10\%}  & 0.2 & $0.995\pm 0.003$  &  $-7.6 \pm 0.5$ & 91.9~\GeV \\ \hline
\mbox{60--80\%} & 0.2 & $0.989 \pm 0.002$ &  $-6.0 \pm 0.3$ & 92.9~\GeV \\ \hline
\mbox{0--10\%}  & 0.4 & $1.027\pm 0.004$  &  $-17.7 \pm 0.5$ & 85.0~\GeV \\ \hline
\mbox{60--80\%} & 0.4 & $0.964 \pm 0.002$ &  $-2.3 \pm 0.2$ & 94.1~\GeV \\ \hline
\end{tabular}
}
\caption{
 The relationship between the mean truth-jet transverse momenta, \avgpttrue, 
and corresponding reconstructed jet transverse momenta, \pTrec. Sample values 
of $\alpha$ and $\beta$ obtained from linear fits to $\avgpttrue(\pTrec)$ 
(see text) according to Eq.~(\ref{eq:avgpttruedef})
and the resulting \avgpttrue\ for $\pTrec = 100$ GeV.
 }
\label{tbl:avgpttrue}
\end{table*}

MC studies indicate that the efficiency for 
PYTHIA jets to be 
reconstructed and to pass UE jet rejection exceeds 98\% for $\pTjet > 60$~\GeV\ in 
the \mbox{0--10\%} centrality bin. 
  For kinematic selection of jets used in this 
study, the jet reconstruction was fully efficient.

\begin{figure}[t]
\centerline{
\includegraphics[width=1.1\fighalfwidth]{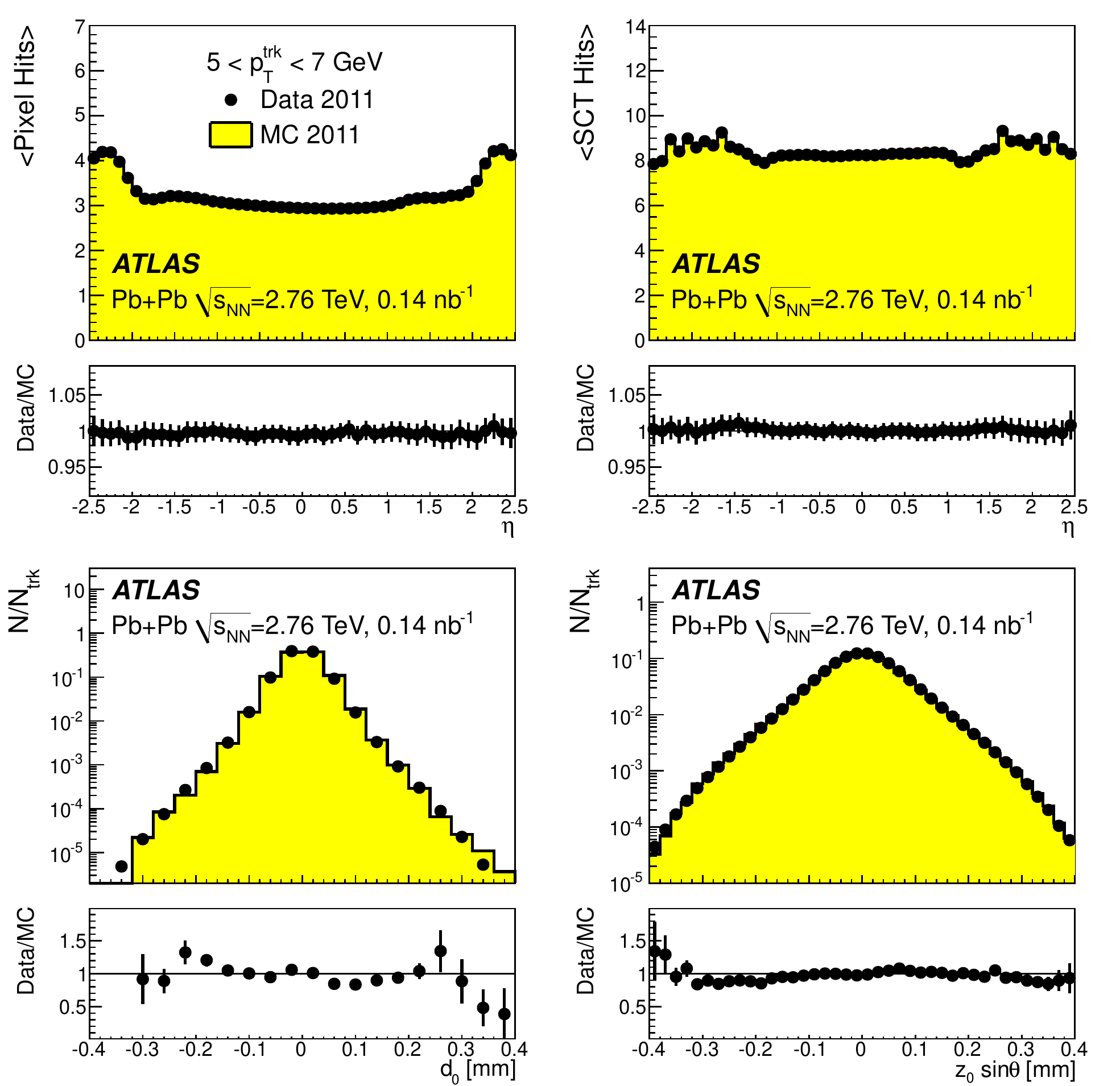}
}
\caption{
  Comparison between data and MC distributions for four different charged-particle 
reconstruction selection parameters. The distributions are shown for 
the \mbox{0--10\%} centrality bin and for charged-particle transverse momenta in the 
range $5 < \pTch < 7$~\GeV. Top: average number of pixel (left) and SCT 
(right) hits per track. 
  Bottom: distribution of track impact parameters with respect to 
the reconstructed primary vertex; both transverse, $d_0$ (left), and 
longitudinal, $z_0 \sin\theta$ (right), impact parameters are shown.
  Ratios of distributions in data to those in MC simulation are shown for each quantity.
 }
\label{fig:trkperf}
\end{figure}
  The efficiency for reconstructing charged particles within jets in \PbPb\ collisions 
was evaluated using the MC sample. Fig.~\ref{fig:trkperf} shows comparisons of 
distributions of four important track-quality variables between data and MC simulation 
for reconstructed tracks over a narrow \pTch\ interval, $5 < \pTch < 7$~\GeV, to 
minimize the impact of differences in MC and data charged-particle \pTch\ 
distributions. The ratios of the data to MC distributions also shown in the figure 
indicate better than 1\% agreement in the $\eta$ dependence of the average number of 
pixel and SCT hits associated with the tracks. The distributions of $d_0$ and $z_0\sin{\theta}$ 
agree to $\lesssim 10\%$ except in the tails of the distributions, which contribute a 
negligible fraction of the distribution. For the purpose of evaluating the track 
reconstruction performance and for the evaluation of response matrices that are used in 
the unfolding (described below),
  the reference ``truth'' particles were taken from the set of final-state 
PYTHIA charged particles.
  These were matched to
reconstructed charged particles using associations between detector
hits and truth tracks recorded by the ATLAS Geant4 simulations. Truth
particles for which no matching reconstructed particle was found were
considered lost due to inefficiency.

\begin{figure}[t]
\centerline{
\includegraphics[width=0.9\fighalfwidth]{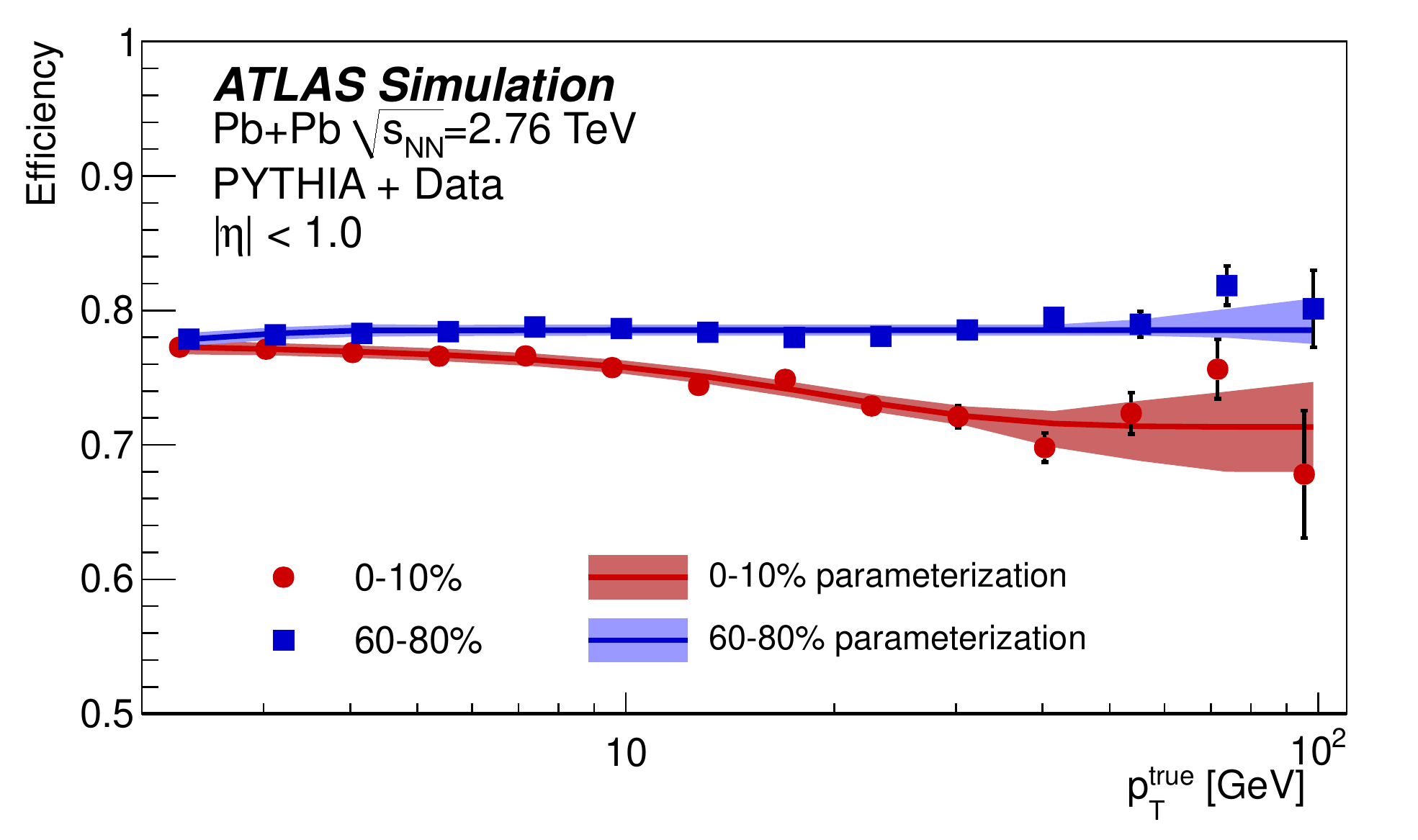}
}
\centerline{
\includegraphics[width=0.9\fighalfwidth]{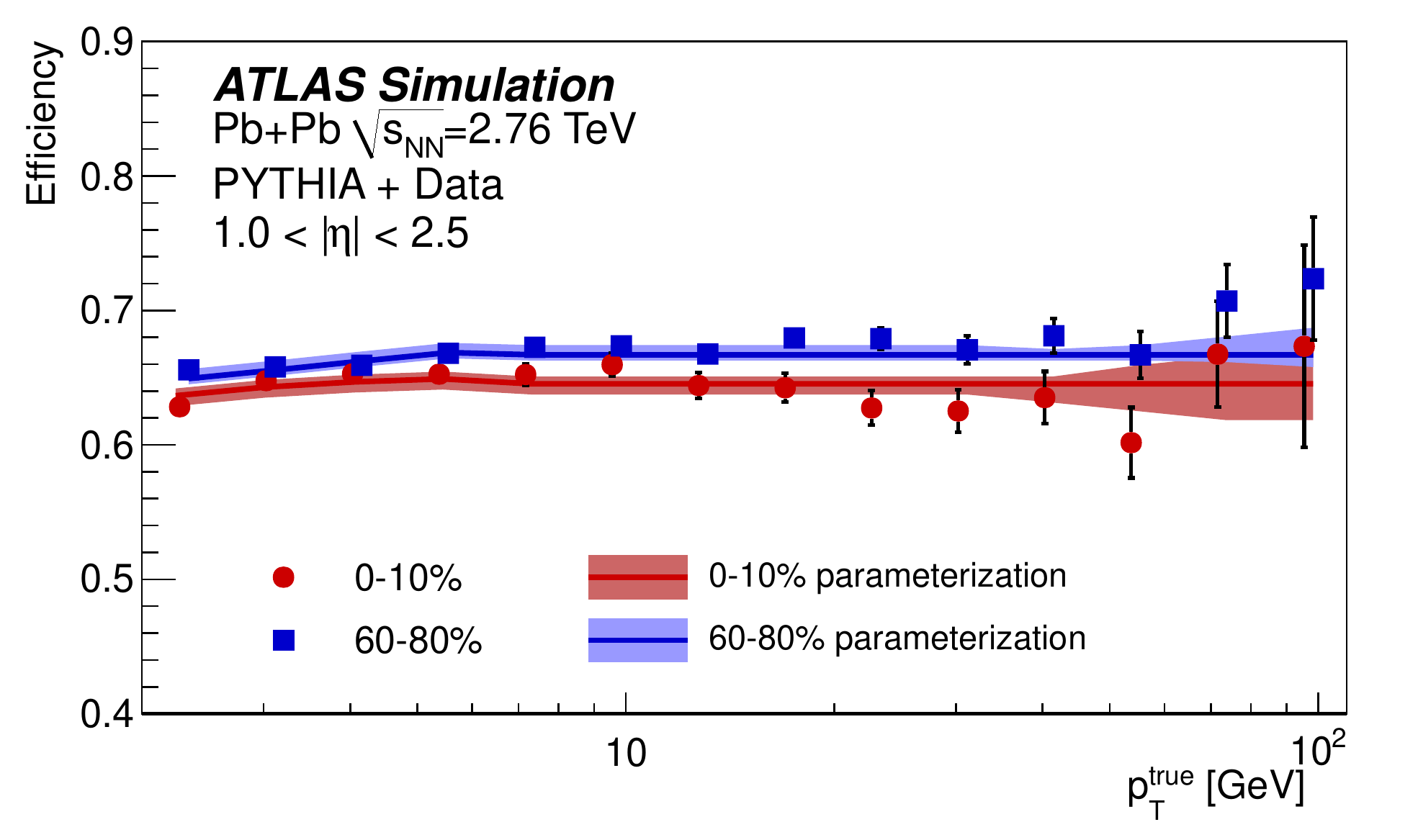}
}
\vspace{-0.15in}
\caption{
  Charged-particle reconstruction efficiency as a function of truth \pt, for 
\mbox{0--10\%} (red) and \mbox{60--80\%} (blue) centrality bins in the region $|\eta| < 1$
(top) and $1<|\eta|<2.5$ (bottom). The 
\pt\ values for the \mbox{0--10\%} points are shifted for clarity. The solid curves 
show parameterizations of efficiencies. The shaded bands show the systematic 
uncertainty in the parameterized efficiencies (see text).
  }
\label{fig:effic}
\end{figure}

The charged-particle
reconstruction efficiency, \effpteta, was evaluated separately in
each of the seven centrality bins used in this analysis for truth 
particles within $\Delta R = 0.4$ of \RFour\ truth jets having
$\pTtrue > 100$~\GeV. Fig.~\ref{fig:effic} shows the efficiency as a
function of truth-particle \pT\ averaged over
$|\eta|<1$ (top) and $1 < |\eta| < 2.5$ (bottom) for the \mbox{0--10\%} and
\mbox{60--80\%} centrality bins. For $\pT < 8$~\GeV, \effpteta\ was
directly evaluated using fine bins in \pT\ and $\eta$. 
  For $\pt > 8$~\GeV\ the \pt\ dependence of the efficiencies were
parameterized separately in the two pseudorapidity intervals shown in 
Fig.~\ref{fig:effic} using a functional form that describes trends at 
low \pt\ as well as at high \pt.
  An example of the resulting parameterizations
is shown by the solid curves in Fig.~\ref{fig:effic}. A
centrality-dependent systematic uncertainty in the parameterized
efficiencies, shown by the shaded bands in Fig.~\ref{fig:effic}, was
evaluated based on both the uncertainties in the 
parameterization and on observed variations of the efficiency with
\pt, which largely result from loss of hits in the SCT at higher
detector occupancy. Thus, the systematic uncertainty in the \mbox{60--80\%}
centrality bin is small because no significant variation of the
efficiency is observed at low detector occupancy, while the
uncertainties are largest for the \mbox{0--10\%} centrality bin with the
largest detector occupancies.

The efficiencies shown in Fig.~\ref{fig:effic}
decrease by about 12\% between the $|\eta| < 1$ interval covered by
the SCT barrel and the $1 < |\eta|< 2.5$ interval covered primarily by
the SCT end-cap. More significant localized drops in efficiency of
about 20\% are observed over $1 < |\eta| < 1.2$ and $2.3 < |\eta| < 2.5$
corresponding to the transition between the SCT barrel and end-cap and
the detector edge respectively. To account for this
and other localized variations of the high \pt\ reconstruction
efficiency with pseudorapidity, the parameterizations in
Fig.~\ref{fig:effic} for $\pt > 8$~\GeV\ are multiplied by an
$\eta$-dependent factor evaluated in intervals of 0.1~units to produce
\effpteta.  

\section{Fragmentation functions and unfolding}
Jets used for the fragmentation measurements presented here were
required to have $\pTjet > 85, 92$ and 100~\GeV for $R =
0.2, 0.3,$ and 0.4 jets, respectively. The jet thresholds for \RThree\ and
\RTwo\ jets represent the typical energy measured with the smaller jet
radii for an \RFour\ jet with $\pt = 100$~\GeV. Jets were also
required to have either $ 0 < |\eta| < 1$ or $1.2 < |\eta| < 1.9$. The
restriction of the measurement to $|\eta| < 1.9$ avoids the region at
the detector edge with reduced efficiency ($|\eta|>2.3$). 
    The exclusion of the range
  $1 < |\eta|< 1.2$ removes from the measurement jets whose large-$z$
  fragments, which are typically collinear with the jet axis, would be
  detected in the lower-efficiency $\eta$ region spanning the gap
  between SCT barrel and end-cap. While this exclusion does not
  significantly change the result of the measurement, it reduces the
  systematic uncertainties at large $z$ or \pTch. 

The fragmentation functions were
measured for charged particles with $\pTch > 2$~\GeV\ within an angular
range $\Delta R = 0.4$ of the jet direction for all three $R$ values used in the jet
reconstruction. To reduce the effects of the UE broadening of
the jet position measurement, for \RThree\ and \RFour\ jets, the jet
direction was taken from that of the closest matching \RTwo\ jet
within $\Delta R = 0.3$ when such a matching jet was found. For each  
charged particle, the longitudinal jet momentum fraction, $z$, was
calculated according to 
\begin{equation}
z  =  \frac{\pTch}{\pTjet} \cos \Delta R,
\end{equation}
where $\Delta R$ here represents the angle between the charged
particle and jet directions.\footnote{The $\Delta R$ is a boost-invariant replacement for the polar angle
$\theta$.}

Charged particles from the UE contribute a \pTch- and
centrality-dependent background to the measurement
that must be subtracted to obtain the true 
fragmentation functions. The contribution of the UE background was 
separately evaluated for $R = 0.2, 0.3,$ and 0.4 jets in events
having at least one such jet above the jet \pT\ thresholds
using a grid of $\Delta R = 0.4$ cones that spanned
the full coverage of the inner detector. Any such cone having a
charged particle with $\pTch > 6$~\GeV\ was assumed to be associated
with a real jet in the event and was excluded from the UE background
determination. 
  The threshold of 6 GeV was chosen to be high enough to avoid bias of the UE \pTch\ 
distribution.

The resulting per-jet UE charged-particle yields, $\mathrm{d}\nchUE/\mathrm{d}\pTch$ 
were evaluated over $2 < \pTch < 6$~\GeV\ as a function of
\pTch, \pTjet, and \etajet, averaged over all cones
in all events within a given centrality bin according to:
\begin{equation}
\dfrac{\mathrm{d} \nchUE}{\mathrm{d} \pTch} = 
\dfrac{1}{\Ncone}\dfrac{\Delta N_{\mathrm{ch}}^{\mathrm{cone}}(\pTch,\pTjet,
  \etajet)}{\Delta \pTch}. 
\label{eq:background}
\end{equation}
  Here \Ncone\ represents the number of background cones 
  having a jet of a given radius above the
corresponding \pTjet\ threshold, and $\Delta N_{\mathrm{ch}}^{\mathrm{cone}}$
represents the number of charged particles in a given \pTch\ bin in
all such cones evaluated for jets with a given \pTjet\ and
\etajet. Not shown in Eq.~(\ref{eq:background}) is a correction factor
that was applied to each background cone to correct for the difference
in the average UE-particle yield at a given \pTch\ between the  $\eta$
position of the cone and \etajet, and a separate correction factor to account
for the difference in the elliptic flow modulation at the $\phi$ position of
the UE cone and \phijet. That correction was based on a
parameterization of the \pTch\ and centrality dependence of previously
measured elliptic flow coefficients, $v_2$ \cite{ATLAS:2011ah}.

\begin{figure*}
\centerline{
\includegraphics[width=\textwidth]{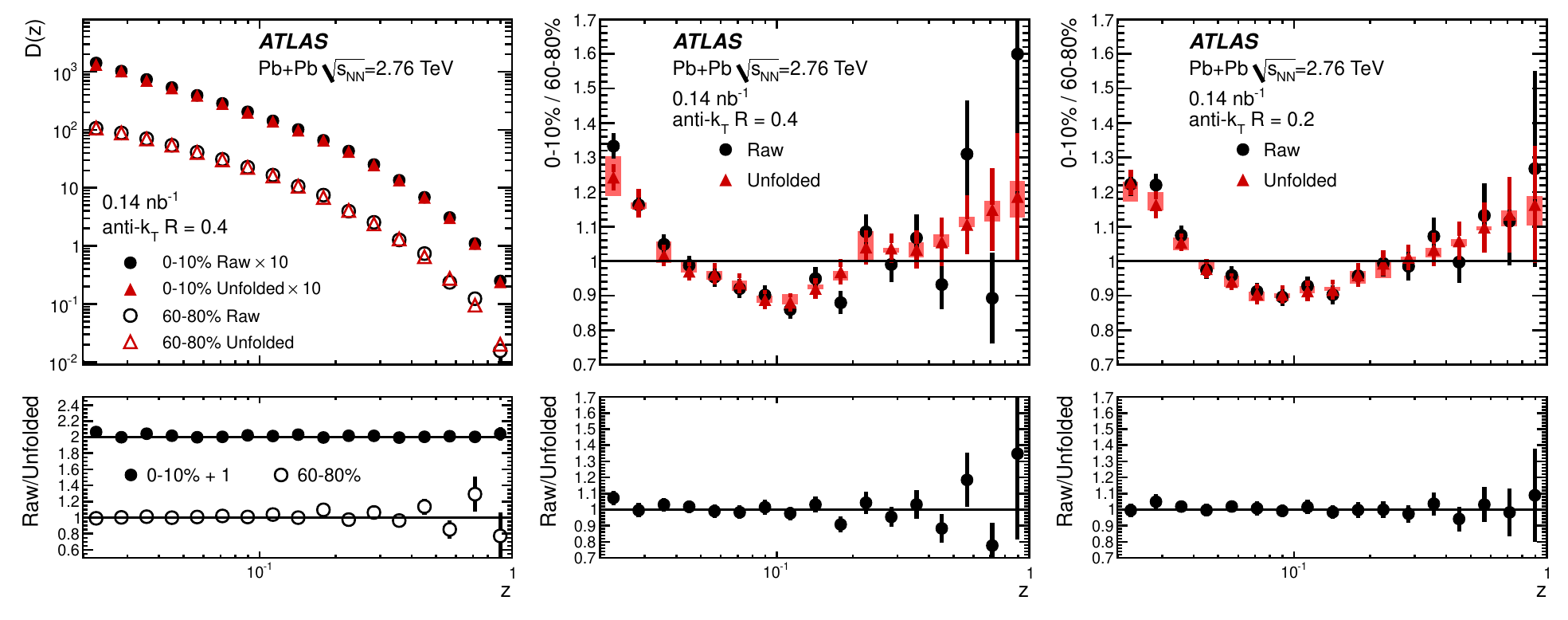}
}
\vspace{-0.15in}
\caption{
Measured and unfolded \Dz\ distributions for \RFour\ and \RTwo\ jets in central
(\mbox{0--10\%}) and peripheral (\mbox{60--80\%}) collisions. Top left: \RFour\ \Dzmeas\ and
\Dz\ distributions, bottom left: ratios of measured to unfolded
\RFour\ \Dz\ distributions with the \mbox{0--10\%} shifted by $+1$ for clarity.
Top middle and right: central-to-peripheral ratios of 
measured (\Rdzmeas) and unfolded (\Rdz) distributions for \RFour\ and
\RTwo, respectively. Bottom middle and right: 
ratio of \Rdzmeas\ to \Rdz\ for \RFour\ and \RTwo, respectively.
}
\label{fig:fragdemo}
\end{figure*}
By evaluating the UE contribution only from events containing jets
included in the analysis, the background automatically has the correct
distribution of centralities within a given centrality bin.  The
$\mathrm{d}\nchUE/\mathrm{d}\pTch$ is observed to be independent of \pTjet\ both in the
data and MC simulation. That observation excludes the possibility that the
upfeeding of jets in \pTjet\ due to the finite JER could induce a
dependence of the UE on jet \pT. However, such upfeeding was observed
to induce in the MC events a \pTjet-independent, but centrality-dependent
mismatch between the extracted $\mathrm{d}\nchUE/\mathrm{d}\pTch$ and the actual UE
contribution to reconstructed jets. That mismatch was found to result
from intrinsic correlations between the charged-particle density in
the UE and the MC \pTjet\ error, $\Delta
p_{\mathrm{T}}^{\mathrm{jet}} = \pTrec - \pTtrue$.  In particular, jets with positive
(negative) $\Delta p_{\mathrm{T}}^{\mathrm{jet}}$ are found to have an
UE contribution larger (smaller) than jets with $\Delta
p_{\mathrm{T}}^{\mathrm{jet}} \sim 0$.  Due to the net upfeeding on
the falling jet spectrum, the selection of jets above a given
\pTjet\ threshold causes the UE contribution to be larger than that
estimated from the above-described procedure. The average fractional
mismatch in the estimated UE background was found to be independent of
\pTch\ and to vary with centrality by factors between 1.04-1.08, 1.07-1.10, and 1.12-1.15
for $R = 0.2, 0.3,$ and 0.4, respectively. The measured
$\mathrm{d}\nchUE/\mathrm{d}\pTch$ values in the data were corrected by these same
factors before being subtracted.

Two different sets of charged-particle fragmentation distributions were
measured for each centrality bin and $R$ value:
 \begin{equation}
\Dptmeas \equiv \frac{1}{\varepsilon} \left(\frac{1}{\Njet} 
\frac{\Delta \Nch}{\Delta \pTch} - \frac{\mathrm{d} \nchUE}{\mathrm{d} \pT}\right), \\
\end{equation}
and 
 \begin{equation}
\Dzmeas \equiv \frac{1}{\varepsilon} \left(\frac{1}{\Njet} 
\frac{\Delta \Nch}{\Delta z} - \left.\frac{\mathrm{d} \nchUE}{\mathrm{d} \pT}\right|_{\pTch=z\pTjet}\right),
\end{equation}
where \Njet\ represents the total number of jets passing the
above-described selection cuts in a 
given centrality bin, and $\Delta \Nch$ 
represents the number of measured charged particles within $\Delta R =
0.4$ of the jets in given bins of \pTch\ and $z$, respectively. 
The efficiency correction, $1/\varepsilon$, was applied on a per-particle basis
using the parameterized MC efficiency, \effpteta, assuming $\pTchtrue = \pTchrec$.  While
that assumption is not strictly valid, the efficiency varies
sufficiently slowly with \pTchtrue\ that the error introduced by this
assumption is $\lesssim 1\%$ everywhere.

\begin{figure*}
\centerline{
\includegraphics[width=\textwidth]{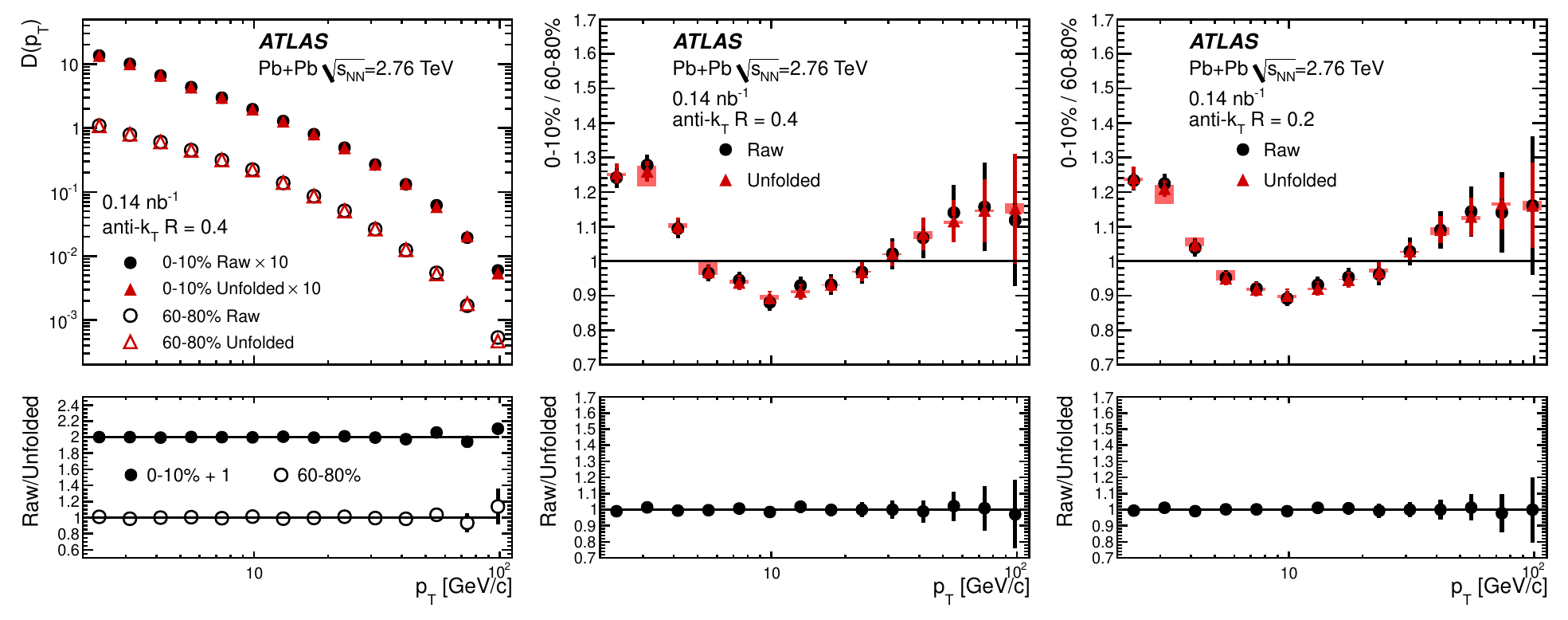}
}
\vspace{-0.15in}
\caption{
Measured and unfolded \Dpt\ distributions for \RFour\ and \RTwo\ jets in central
(\mbox{0--10\%}) and peripheral (\mbox{60--80\%}) collisions. Top left: \RFour\ \Dptmeas\ and
\Dpt\ distributions, bottom left: ratios of measured to unfolded
\RFour\ \Dpt\ distributions with the \mbox{0--10\%} shifted by $+1$ for clarity.
Top middle and right: central-to-peripheral ratios of 
measured (\Rdptmeas) and unfolded (\Rdpt) distributions for \RFour\ and
\RTwo, respectively. Bottom middle and right: 
ratio of \Rdptmeas\ to \Rdpt\ for \RFour\ and \RTwo, respectively.
}
\label{fig:fragdemotwo}
\end{figure*}

The measured \Dzmeas\ distributions for \RFour\ jets in the \mbox{0--10\%} and
\mbox{60--80\%} centrality bins are shown in the top left panel in
Fig.~\ref{fig:fragdemo}.  The top middle panel shows the ratio of
\Dzmeas\ between central (\mbox{0--10\%}) and peripheral (\mbox{60--80\%}) collisions,
$\Rdzmeas \equiv \Dzmeas|_{0-10}/\Dzmeas|_{60-80}$. For comparison,
the \Dzmeas\ ratio is shown on the top right panel for \RTwo\ jets.
Similar plots are shown in Fig.~\ref{fig:fragdemotwo} but for
\Dptmeas. The \Dzmeas\ ratios for both \RTwo\ and \RFour\ indicate an
enhanced fragment yield at low $z$, $z \lesssim 0.04$, in jets in the
\mbox{0--10\%} centrality bin compared to jets in the \mbox{60--80\%} centrality bin
and a suppressed yield of fragments with $z \sim 0.1$. Similar results
are observed in the \Dptmeas\ ratios over the corresponding
\pT\ ranges. The \RTwo\ \Dzmeas\ and the \RTwo\ and \RFour\ \Dptmeas\ ratios 
rise above one for $z \gtrsim 0.2$ or $\pT \gtrsim 25$~\GeV. However,
the ratios differ from one by only $\text{1--2}\sigma
(\mathrm{stat})$. No such variations of the \Dzmeas\ and
\Dptmeas\ distributions with centrality as seen in the data 
are observed in the MC simulation. The central-to-peripheral 
ratios of MC \Dzmeas\ and \Dptmeas\ distributions for
\RFour\ and \RTwo\ jets (not shown) are within 3\% of one for all $z$ and \pT.

The \Dptmeas\ and \Dzmeas\ distributions were unfolded using 
a one-dimensional Singular Value Decomposition (SVD) method
\cite{Hocker:1995kb} implemented in {\tt RooUnfold} \cite{roounfold}
to remove the effects of charged particle and jet \pT\ resolution. The
SVD method implements a regularized matrix-based unfolding that attempts to
``invert'' the equation {\boldmath{$b = Ax$}}, where
{\boldmath{$x$}}, is a true spectrum, {\boldmath{$b$}} is an observed
spectrum, and {\boldmath $A$} is the ``response matrix'' that describes
the transformation of {\boldmath$x$} to {\boldmath $b$}. 
For \Dpt, the unfolding
accounts only for the charged-particle \pT\ resolution and uses a
response matrix derived from the MC data set that describes the
distribution of reconstructed \pTch\  as a function of MC truth
\pTch. The response matrix {\boldmath$A$}$(\pTchrec, \pTchtrue)$ is
filled using the procedures described in Section~\ref{sec:perform}. The
\Dz\ unfolding simultaneously accounts for both charged particle and
jet resolution using a response matrix {\boldmath  $A$}$(\zrec,
\ztrue)$ with \ztrue\ (\zrec) calculated using purely truth
(fully reconstructed) quantities. A cross-check was performed for the
\Dz\ unfolding that included only the jet energy resolution to ensure
that the combination of the two sources of resolution in the
one-dimensional unfolding did not distort the result. Because the
\Dzmeas\ and \Dptmeas\ distributions were already corrected for the
charged-particle reconstruction efficiency, the response matrices were
only populated with truth particles for which a reconstructed particle
was obtained and each entry was corrected for reconstruction
efficiency so as to not distort the shape of the true distributions.

To ensure that 
statistical fluctuations in the MC \pTtrue\ or \ztrue\ distributions
do not distort the unfolding, those distributions
were smoothed by fitting them to appropriate functional forms. The
truth \Dpt\ distributions were fit to polynomials in $\ln(\pT)$. The
truth \Dz\ distributions were parameterized using an extension of a
standard functional form \cite{FragFit}, 
\begin{equation}
D(z) = a \cdot z^{d_1}(1+c-z)^{d_2} \cdot (1+b \cdot (1-z)^{d_3}), 
\end{equation}
where $a, b, c, d_{i}$ were free parameters of the fit. The
non-standard additional parameter ``$c$'' was added to improve the
description of the truth distribution at large $z$. When filling the
truth spectra and response matrices, the entries were
weighted to match the truth spectra to the fit functions.

The SVD unfolding was performed using a regularization parameter
obtained from the ninth singular value ($k =9$) of the unfolding matrix. 
Systematic uncertainties in the unfolding due to regularization were
evaluated by varying $k$ over the range 5--12 for which the unfolding
was observed to be neither significantly biased by regularization nor unstable.
The statistical uncertainties in the unfolded spectra were obtained
using the pseudo-experiment method \cite{Hocker:1995kb}. The largest
absolute uncertainty obtained over $5 \leq k \leq 12$  was taken
to be the statistical uncertainty in the unfolded result.

Unfolded fragmentation functions, \Dz, are shown in the top left panel
in Fig.~\ref{fig:fragdemo} and compared to the corresponding
\Dzmeas\ distributions for \RFour\ jets in central (\mbox{0--10\%})
and peripheral (\mbox{60--80\%}) collisions. Similar results for \Dpt\ are
shown in Fig.~\ref{fig:fragdemotwo}. For both figures, the ratios of unfolded to measured
distributions are shown in the bottom left panel with the ratio for
\mbox{0--10\%} centrality bin offset by $+1$. Those ratios show that the
unfolding has minimal impact on the fragmentation functions in
both peripheral and central collisions. Only the largest $z$ point in
the \mbox{0--10\%} bin changes by more than 20\%. 

The middle and top right panels in Fig.~\ref{fig:fragdemo}
(Fig.~\ref{fig:fragdemotwo}) show for \RFour\ and \RTwo\ jets, respectively 
the ratios of unfolded \Dz\ (\Dpt) distributions, 
$\Rdz \equiv \Dz|_{0-10}/\Dz|_{60-80}$ ($\Rdpt \equiv
\Dpt|_{0-10}/\Dpt|_{60-80}$), compared to the ratios before
unfolding. The unfolding reduces the \Dz\ ratio slightly at low $z$ but
otherwise leaves the shapes unchanged. To evaluate
the impact of the unfolding on the difference between central and
peripheral fragmentation functions, the middle and bottom right panels in
Fig.~\ref{fig:fragdemo} (Fig.~\ref{fig:fragdemotwo}) show the ratio
of \Rdzmeas\ (\Rdptmeas) to \Rdz (\Rdpt). Except
for the lowest $z$ point, the ratio is consistent with one over the
entire $z$ range. Thus, the features observed in
\Rdzmeas\ (\Rdptmeas), namely the 
enhancement at low $z$ (\pT) in central collisions relative to peripheral
collisions, the suppression at intermediate $z$ (\pT), and the rise above
one at large $z$ (\pT) are robust
with respect to the effects of the charged particle and jet
\pT\ resolution. 

\section{Systematic uncertainties}
\label{sec:systerr}

Systematic uncertainties in the unfolded \Dz\ and \Dpt\ distributions
can arise due to uncertainties in the jet energy scale and jet energy
resolution, from systematic uncertainties in the unfolding procedure including
uncertainties in the shape of the truth distributions, uncertainties
in the charged particle reconstruction, and from
the UE subtraction procedure.

The systematic uncertainty due to the jet energy scale (JES) has two contributions, 
an absolute JES uncertainty and an uncertainty in the variation of the JES from 
peripheral to more central collisions. 
  The absolute JES uncertainty was determined by shifting the transverse momentum 
of the reconstructed jets according to the evaluation of the jet energy scale 
uncertainty in Ref.~\cite{Aad:2011he}.
  The typical size of the JES uncertainty for jets used in this study is 2\%.
  The shift in the JES has negligible impact on the ratios 
between central and peripheral events 
  of \Dpt\ and \Dz\ 
distributions whereas it has a clear impact on the \Dpt\ and \Dz\ 
distributions. At high \pt\ or $z$ the resulting uncertainty reaches 15\%.
  The evaluation of centrality-dependent uncertainty on JES uses the 
estimates from Ref.~\cite{jetsupp}. The centrality-dependent JES uncertainty 
is largest for the most central collisions where it reaches 1.5\%.
  The evaluation of the jet energy resolution (JER) uncertainty follows the 
procedure applied in proton--proton jet measurements 
 \cite{Aad:2012ag}.
The typical size of JER uncertainty for jets used in the study is less than 
2\%.
  This uncertainty is centrality independent since the dijets in MC are overlayed to real data.
  The resulting combined systematic uncertainty from JER and centrality-dependent 
JES on the ratios reaches 6\% at high \pt\ and 10\% at high $z$ and 
it has a similar size in the case of \Dpt\ or \Dz\ distributions as in the 
case of their ratios.

The systematic uncertainty associated with the unfolding is connected
with the sensitivity of the unfolding procedure to the choice of
regularization parameter and to the parameterization of the truth
distribution.  
  The uncertainty due to the choice of regularization parameter 
was evaluated by varying $k$ over the range 5--12.
  The typical systematic uncertainty is found to be smaller than 3\% or 2\% 
for the \Dz\ or \Dpt, respectively.  The systematic uncertainty due to the 
parameterization of the truth distribution was determined from the 
statistical uncertainties of the fits to these distributions. This systematic 
uncertainty is below 1\% or 2\% for the \Dz\ or \Dpt, respectively.

\begin{figure*}
\centerline{
\includegraphics[height=4.5in]{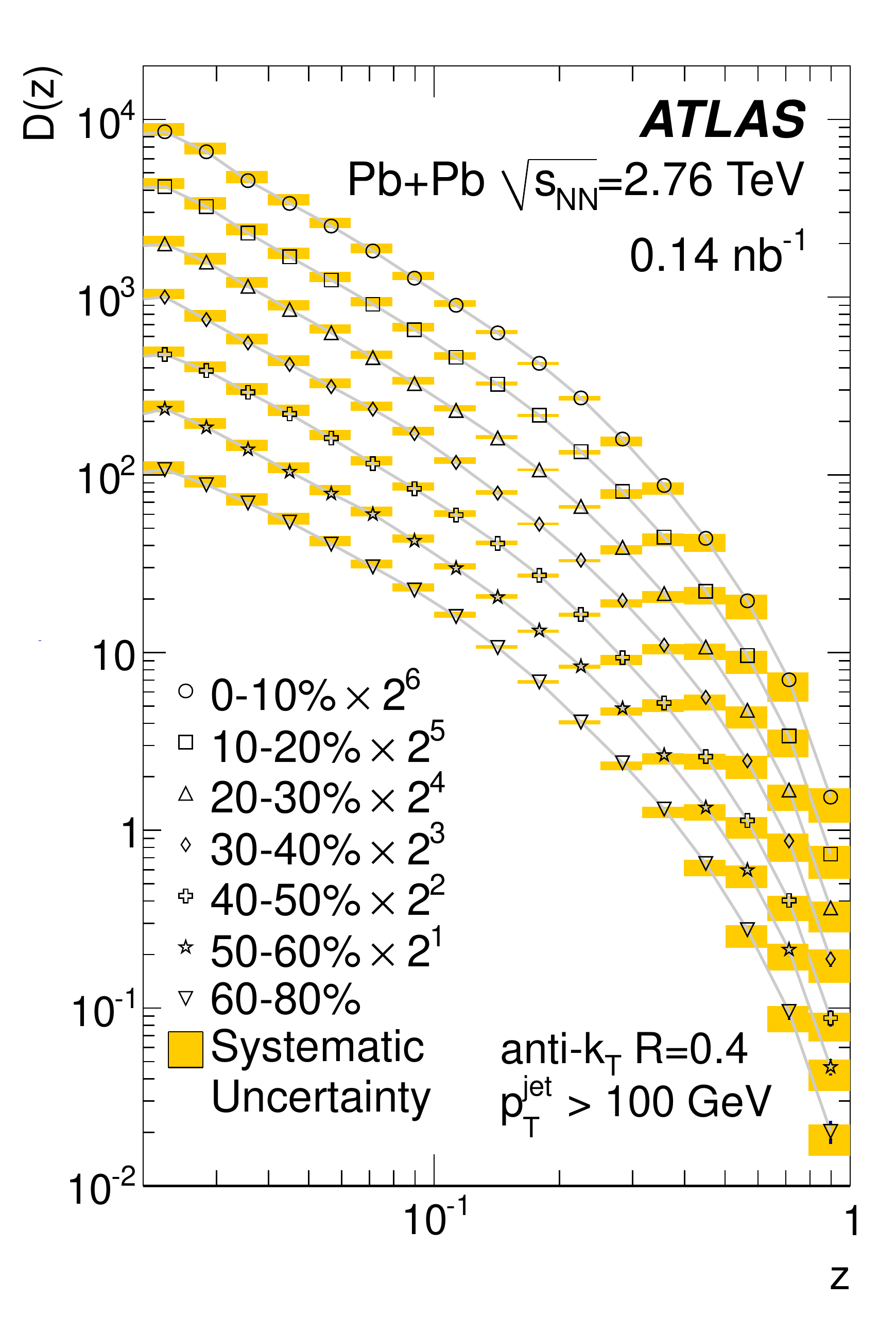}		
\includegraphics[height=4.5in]{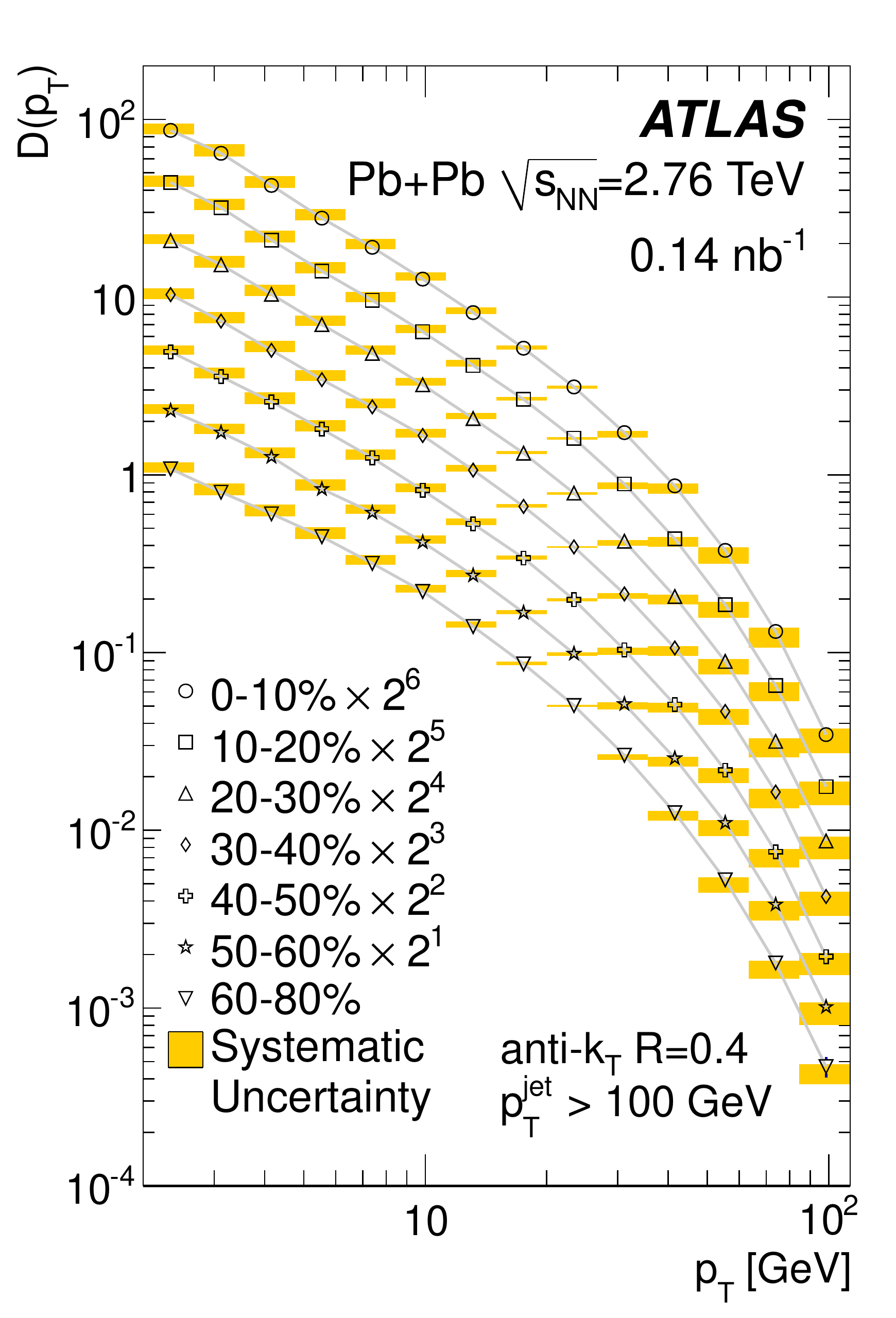}
}
\vspace{-0.15in}
\caption{
  Unfolded \RFour\ longitudinal charged particle fragmentation function, \Dz\ and the charged particle 
transverse momentum distribution, \Dpt, for the seven centrality bins included in this analysis. The 
statistical uncertainties are everywhere smaller than the points. The yellow shaded error bars indicate 
systematic uncertainties. Grey lines connecting the central values of distributions are to guide the eye.
}
\label{fig:dz}
\end{figure*}
  The estimate of systematic uncertainty due to the tracking efficiency follows 
methods of the inclusive charged particle measurement 
  \cite{ATLAS:2011ah}. 
The uncertainty is quantified using the error of the fit of tracking efficiency 
and by varying the tracking selection criteria. In the intermediate-\pt\ region 
the systematic uncertainty is less than 2\%. In the low and high \pt\ region the 
systematic uncertainty is larger, but less than 8\%.

  An independent evaluation of potential systematic uncertainties in the
central-to-peripheral ratios of \Dz\ and \Dpt , due to
all aspects of the analysis, was obtained by evaluating the deviation from unity
of the MC central (\mbox{0--10\%}) to peripheral (\mbox{60--80\%}) ratios of the
fragmentation functions.  
  Since there is no jet quenching employed in MC simulation, the ratios are expected 
not to show any deviation from unity. No deviation from unity is indeed 
observed, the largest localized deviation is $\lesssim 4\%$.
  To quantify the deviations from unity,
the MC \Rdz\ and \Rdpt\ ratios were fit by piecewise continuous
functions composed of linear functions defined over the $z$ (\pT) ranges
$z$=\mbox{0.02--0.06} ($\pt$=\mbox{2--6}~GeV), $z$=\mbox{0.06--0.3} ($\pt$=\mbox{6--30}~GeV), and
$z>0.3$ ($\pt\ > 30$~GeV) with parameters constrained such that the
linear functions match at the boundaries. The resulting fits are used
as estimates of the systematic uncertainties on all measured \Rdz\ and
\Rdpt\ ratios reported in Section~\ref{sec:results}. This systematic
uncertainty is certainly correlated with and may overlap with other
systematic uncertainties described above.

\section{Results}
The unfolded fragmentation functions, \Dz\ and \Dpt, for \RFour\ jets are shown
in Fig.~\ref{fig:dz} for the seven centrality bins included in the
analysis with the distributions for different centralities multiplied
by successive values of two for presentation purposes. 
  The shaded error bands indicate systematic uncertainties as discussed 
in the previous section.
  The \Dpt\ and \Dz\ distributions have similar
shapes that are characteristic of fragmentation functions with a steep
drop at the endpoint. 
\label{sec:results}

\begin{figure*}        
\centerline{
\includegraphics[width=0.90\textwidth]{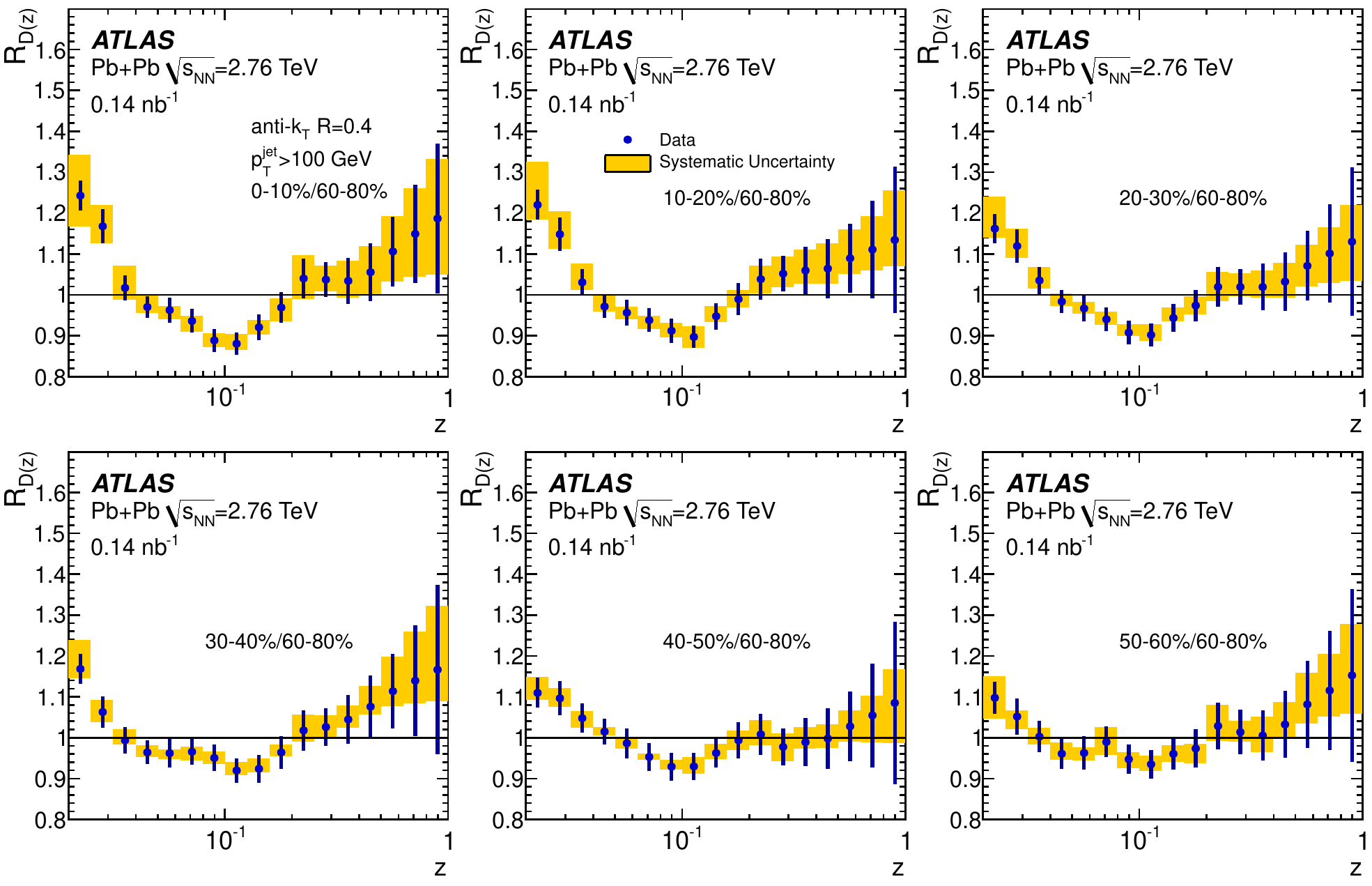}	
}
\vspace{-0.15in}
\caption{
Ratios of \Dz\ for six bins in collision centrality to those in
peripheral (\mbox{60--80\%}) collisions, \Dzratio, for \RFour\ jets. 
  The error bars on the data points indicate statistical uncertainties while the 
yellow shaded bands indicate systematic uncertainties.
}
\label{fig:dzratiorfour}
\end{figure*}

To evaluate the centrality dependence of the fragmentation functions,
ratios were calculated of the \RFour\ \Dz\ distributions for all
centrality bins excluding the peripheral bin to the \Dz\ measured in
the peripheral, \mbox{60--80\%} centrality bin. The results are shown in
Fig.~\ref{fig:dzratiorfour}.  The ratios for all centralities show an
enhanced yield of low $z$ fragments and a suppressed yield of
fragments at intermediate $z$ values in more central collisions
relative to the \mbox{60--80\%} centrality bin. For the \mbox{0--10\%} centrality bin,
the yield of fragments at $z = 0.02$ is enhanced relative to that in
the \mbox{60--80\%} centrality bin by 25\% while the yield at $z = 0.1$ is
suppressed by about 10\%.  The size of the
observed modifications at low, intermediate, and high $z$ decreases
gradually from central to peripheral collisions.

\begin{figure*}
  \centerline{
  \includegraphics[width=0.90\textwidth]{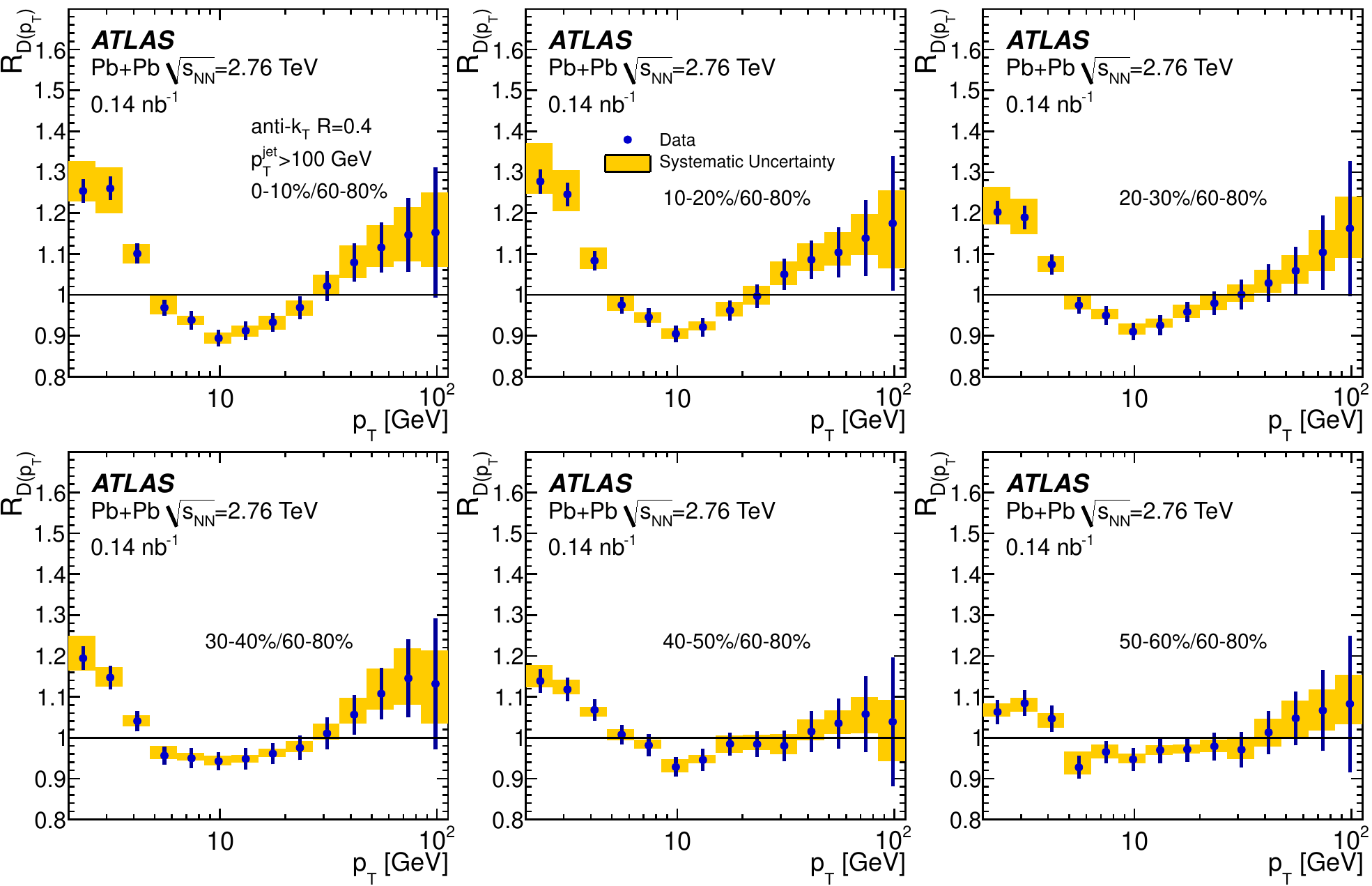} 
  }
  \caption{
  Ratios of unfolded \Dpt\ distributions for six bins in collision centrality 
to those in peripheral (\mbox{60--80\%}) collisions, \Dptratio, for \RFour\ jets.
  The error bars on the data points indicate statistical uncertainties while the 
yellow shaded bands indicate systematic uncertainties.
}
\label{fig:dptratiorfour}
\end{figure*}

The statistical and systematic uncertainties on \Rdz\ grow
as $z \rightarrow 1$ due to the statistical fluctuations on the
\Dz\ distributions at large $z$ and due to the sensitivity of the
steeply falling \Dz\ distributions to JER and JES systematic
uncertainties. The results in Fig.~\ref{fig:dzratiorfour}
show central values for \Rdz\ above one at high $z$ for the \mbox{0--10\%} 
through the \mbox{30--40\%} centrality bins but the \Rdz\ values differ
from one by typically $1\sigma(\mathrm{stat})$.
  Fig.~\ref{fig:dptratiorfour} shows ratios of
\RFour\ \Dpt\ distributions from non-peripheral centrality bins to
those in the peripheral, \mbox{60--80\%} centrality bin. The ratios in the
figure show the same features as the \Dz\ ratios, namely an
enhancement at low \pT, a suppression at intermediate \pT, and an
increase above one at large \pT\ that is  more significant than that
seen for \Dz. The magnitudes of the deviations from one in
the \Dz\ and \Dpt\ ratios are similar in the low, intermediate, and
high $z$ and \pT\ regions.
  This demonstrates that the modifications observed in
Fig.~\ref{fig:dzratiorfour} do not result from distortions of the $z$
measurement due to JER and JES.
\begin{figure*}
\centerline{
\includegraphics[width=0.99\textwidth]{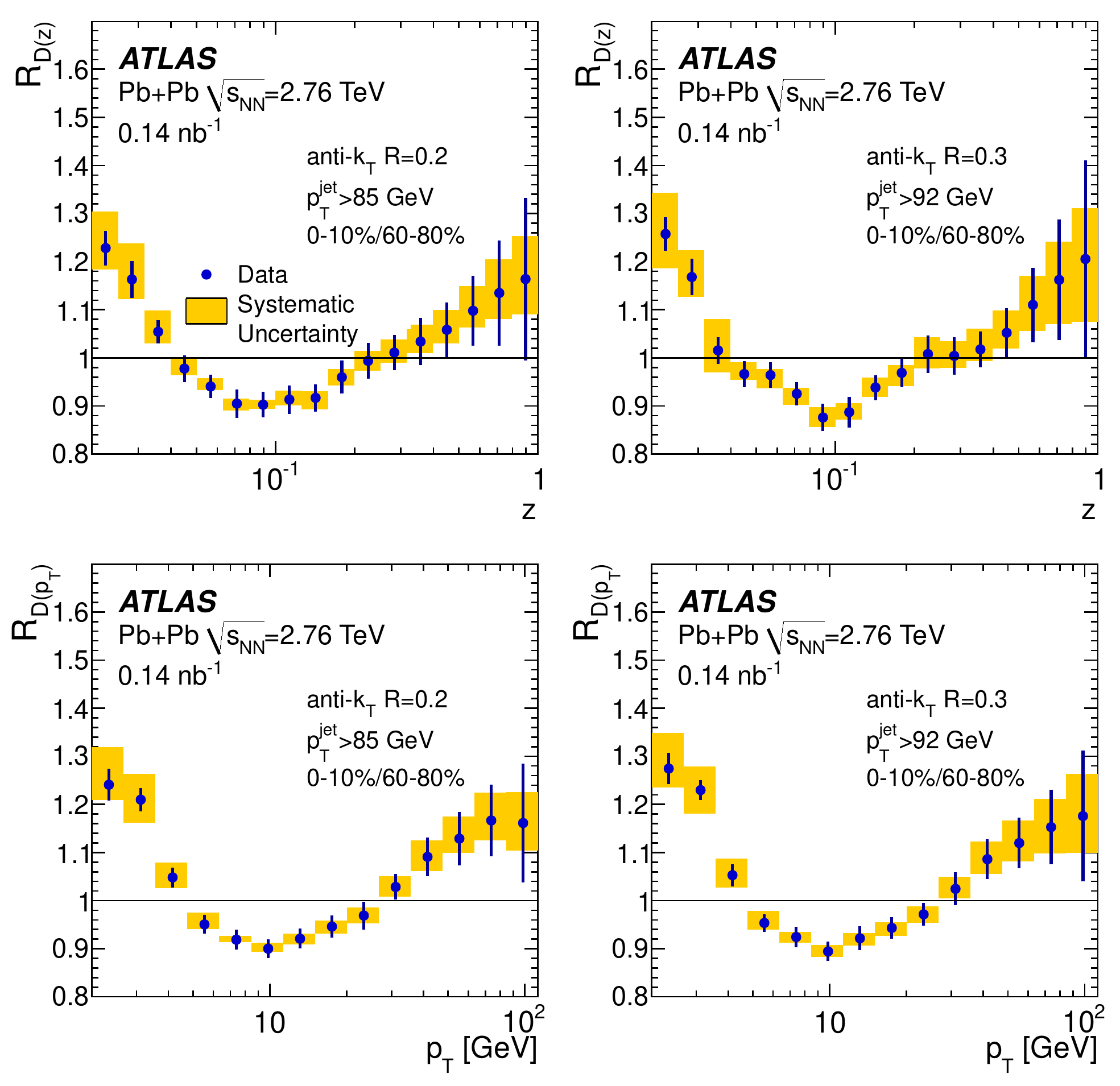}	
}
\caption{
Ratios of unfolded fragmentation functions, \Dz\ (top) and
\Dpt\ (bottom), for central (\mbox{0--10\%}) collisions to those in peripheral
(\mbox{60--80\%}) collisions for \RTwo\ (left) and \RThree\ (right) jets. The
fragmentation functions were evaluated using charged hadrons within
$\Delta R = 0.4$ of the jet axis. 
  The error bars on the data points indicate statistical uncertainties while the 
yellow shaded bands indicate systematic uncertainties.
}
\label{fig:dzratiortwothree}
\end{figure*}

To further demonstrate that the centrality-dependent modifications observed in
\Dz\ and \Dpt\ do not result from unknown UE effects not included in
the systematic uncertainties, Fig.~\ref{fig:dzratiortwothree} shows
ratios of \Dz\ and \Dpt\ distributions between central (\mbox{0--10\%}) and
peripheral (\mbox{60--80\%}) collisions for \RTwo\ and \RThree\ jets.
The fluctuations in the UE are a factor of approximately 100\%
(30\%) smaller for \RTwo\ (\RThree) jets than they are for \RFour\ jets. 
Nonetheless, the features seen in the \RFour\ \Dz\ or \Dpt\ ratios are also
present in the \RTwo\ and \RThree\ ratios with the same
magnitudes. Due to the reduced systematic uncertainties on \Dz\ and
\Dpt\ for \RTwo\ and \RThree\ jets compared to \RFour\ jets, the 
enhancement in the fragmentation functions at large $z$ or \pT\ in
central collisions is more significant for the smaller jet sizes. 

\section{Discussion}
\begin{table*}
\centerline{
{\setlength{\extrarowheight}{6pt}
\begin{tabular}{|r|| r | r | r | r | r | r |} \hline
\multirow{2}{*}{Centrality}  &  \multicolumn{2}{|c|}{$z$=0.02--0.04}  &  \multicolumn{2}{|c|}{$z$=0.04--0.2}  &  \multicolumn{2}{|c|}{$z$=0.4--1.0} \\ \cline{2-7}
                              & $\int \! \Delta D(z) \dz$  & $\int \! z \Delta D(z) \dz$ &$\int \! \Delta D(z) \dz$  & $\int \! z \Delta D(z) \dz$  &
 $\int \! \Delta D(z) \dz$  & $\int \! z \Delta D(z) \dz$ \\ \hline
0-10\% &$  0.79 ^{+0.19}_{-0.25}  $&$  0.020 ^{+0.005}_{-0.007}  $&$ -1.7 ^{+0.6}_{-0.8}  $&
 $  -0.14 ^{+0.04}_{-0.06}  $&$  0.06 ^{+0.05}_{-0.04}  $&$  0.033 ^{+0.026}_{-0.021}$  \\ \hline
10-20\% &$  0.66 ^{+0.17}_{-0.18}  $&$  0.016 ^{+0.005}_{-0.005}  $&$ -1.6 ^{+0.7}_{-0.8}  $&
 $  -0.12 ^{+0.05}_{-0.06}  $&$  0.05 ^{+0.05}_{-0.04}  $&$  0.029 ^{+0.026}_{-0.021}$  \\ \hline
20-30\% &$  0.52 ^{+0.13}_{-0.18}  $&$  0.013 ^{+0.004}_{-0.005}  $&$ -1.3 ^{+0.6}_{-0.6}  $&
 $  -0.12 ^{+0.04}_{-0.04}  $&$  0.04 ^{+0.04}_{-0.04}  $&$  0.025 ^{+0.024}_{-0.020}$  \\ \hline
30-40\% &$  0.39 ^{+0.12}_{-0.17}  $&$  0.009 ^{+0.004}_{-0.005}  $&$ -1.3 ^{+0.6}_{-0.7}  $&
 $  -0.10 ^{+0.04}_{-0.05}  $&$  0.06 ^{+0.04}_{-0.04}  $&$  0.036 ^{+0.020}_{-0.019}$  \\ \hline
40-50\% &$  0.38 ^{+0.11}_{-0.15}  $&$  0.009 ^{+0.003}_{-0.004}  $&$ -0.6 ^{+0.6}_{-0.8}  $&
 $  -0.07 ^{+0.04}_{-0.06}  $&$  -0.01 ^{+0.04}_{-0.04}  $&$  -0.005 ^{+0.024}_{-0.021}$  \\ \hline
50-60\% &$  0.28 ^{+0.15}_{-0.21}  $&$  0.006 ^{+0.004}_{-0.006}  $&$ -1.2 ^{+0.9}_{-0.7}  $&
 $  -0.08 ^{+0.06}_{-0.06}  $&$  0.04 ^{+0.04}_{-0.04}  $&$  0.025 ^{+0.021}_{-0.021}$  \\ \hline
\end{tabular}
}
}
\caption{
  Differences of $D(z)$ distributions in different centralities with respect to peripheral events for \RThree\ jets. The errors represent combined statistical and systematic 
uncertainties.
}
\label{tbl:differences}
\end{table*}

\begin{table*}
\centerline{
{\setlength{\extrarowheight}{6pt}
\begin{tabular}{|r|| r | r | r | r | r | r |} \hline
\multirow{2}{*}{Centrality}  &  \multicolumn{2}{|c|}{$z$=0.02--0.04}  &  \multicolumn{2}{|c|}{$z$=0.04--0.2}  &  \multicolumn{2}{|c|}{$z$=0.4--1.0} \\ \cline{2-7}
                              & $\int \! \Delta D(z) \dz$  & $\int \! z \Delta D(z) \dz$ &$\int \! \Delta D(z) \dz$  & $\int \! z \Delta D(z) \dz$  &
 $\int \! \Delta D(z) \dz$  & $\int \! z \Delta D(z) \dz$ \\ \hline
0-10\% &$  0.65 ^{+0.21}_{-0.20}  $&$  0.017 ^{+0.006}_{-0.005}  $&$ -1.7 ^{+0.5}_{-0.6}  $&
 $  -0.14 ^{+0.04}_{-0.05}  $&$  0.07 ^{+0.05}_{-0.04}  $&$  0.037 ^{+0.030}_{-0.022}$  \\ \hline
10-20\% &$  0.60 ^{+0.16}_{-0.16}  $&$  0.016 ^{+0.005}_{-0.004}  $&$ -1.6 ^{+0.7}_{-0.7}  $&
 $  -0.12 ^{+0.05}_{-0.05}  $&$  0.08 ^{+0.05}_{-0.04}  $&$  0.046 ^{+0.029}_{-0.025}$  \\ \hline
20-30\% &$  0.48 ^{+0.11}_{-0.14}  $&$  0.013 ^{+0.003}_{-0.004}  $&$ -1.6 ^{+0.6}_{-0.5}  $&
 $  -0.13 ^{+0.04}_{-0.04}  $&$  0.04 ^{+0.05}_{-0.04}  $&$  0.026 ^{+0.029}_{-0.024}$  \\ \hline
30-40\% &$  0.44 ^{+0.11}_{-0.15}  $&$  0.011 ^{+0.003}_{-0.004}  $&$ -1.4 ^{+0.6}_{-0.7}  $&
 $  -0.11 ^{+0.05}_{-0.05}  $&$  0.07 ^{+0.04}_{-0.05}  $&$  0.044 ^{+0.021}_{-0.028}$  \\ \hline
40-50\% &$  0.33 ^{+0.09}_{-0.14}  $&$  0.009 ^{+0.003}_{-0.004}  $&$ -1.0 ^{+0.6}_{-0.8}  $&
 $  -0.09 ^{+0.04}_{-0.06}  $&$  -0.03 ^{+0.05}_{-0.04}  $&$  -0.011 ^{+0.030}_{-0.020}$  \\ \hline
50-60\% &$  0.27 ^{+0.12}_{-0.18}  $&$  0.007 ^{+0.003}_{-0.005}  $&$ -1.0 ^{+0.8}_{-0.7}  $&
 $  -0.07 ^{+0.06}_{-0.06}  $&$  0.04 ^{+0.04}_{-0.05}  $&$  0.027 ^{+0.024}_{-0.029}$  \\ \hline
\end{tabular}
}
}
\caption{
  Differences of $D(z)$ distributions in different centralities with respect to peripheral events for \RTwo\ jets. The errors represent combined statistical and systematic 
uncertainties.
}
\label{tbl:differencesR2}
\end{table*}

To quantify the effects of the modifications observed in
Fig.~\ref{fig:dzratiortwothree} on the actual distribution of
fragments within the measured jets, the differences in fragmentation 
functions, $\Delta \Dz = D(z)|_{\mathrm{cent}} -
D(z)|_{\mathrm{60-80}}$ were calculated and integrals of these
distributions, $\int \! \Delta \Dz \dz$
taken over three $z$ ranges chosen to match the observations:
0.02--0.04, 0.04--0.2, and 0.4--1. The last interval was chosen
to focus on the region where $\Rdz > 1$. The results are given in
Table~\ref{tbl:differences} and Table~\ref{tbl:differencesR2} for
\RThree\ and \RTwo\ jets, respectively. Similar results were obtained
for \RFour\ jets but with larger uncertainties. The results presented in the tables indicate an
increase in the number of particles with $0.02 < z < 0.04$
of less than one particle per jet in the \mbox{0--10\%} centrality bin
relative to the \mbox{60--80\%} centrality bin. A decrease of about 1.5
particles per jet is observed for $0.04 < z < 0.2$. 
The differences between the integrals of the fragmentation functions
over $0.4 < z < 1$ are not significant relative to the uncertainties.
The results for $\int \! \Delta \Dz \dz$ shown in the two tables
indicate that in the most central collisions a small fraction, $< 2\%$, of the jet transverse momentum is
carried by the excess particles in  
$0.02 < z < 0.04$ for central collisions, but that the depletion in
fragment yield in $0.04 < z < 0.2$ accounts
on average for about 14\% of \pTjet. 

To better evaluate the significance of the increase in \Rdz\ and
\Rdpt\ above one at large $z$ or \pT, average \Rdz\ and \Rdpt\ ratios
were calculated by summing the central and peripheral \Dz\ or
\Dpt\ distributions over different regions corresponding to the last
$n$ points in the measured distributions, $n = 2-6$.  
  For each resulting average ratio, $\ARdz$ or $\ARdpt$, the significance of the 
deviation from one was evaluated as $(\ARdz - 1)/\SARdz$ or 
$(\ARdpt - 1)/\SARdpt$ where $\sigma$ represents the combined statistical 
and systematic uncertainty.
  Because there is
significant cancellation of systematic uncertainties in the ratios,
this analysis provides a more sensitive evaluation of the significance
of the large-$z$ excess. For \RFour\ jets the combined \Rdz\ (\Rdpt),
differs from one by approximately $1\sigma$ ($1.5\sigma$) for any of the $n$
values. For \RTwo\ jets, \Rdz\ differs from 1 by approximately
$1.5\sigma$ for all $n$ values, while \Rdpt\ differs from one by
$2\sigma$ for $n=3\text{--}6$ corresponding to $\pT >
47.5$~\GeV\ through $\pT > 20$~\GeV. The greater significance of the
deviations of the \RTwo\ \Rdpt\ relative to the \RTwo\ \Rdz\ and the
\RFour\ \Rdz\ and \Rdpt\ can be attributed to the reduced role of the
jet energy resolution in influencing the measurement of the central-to-peripheral ratios for large hadron momenta.

Theoretical predictions for medium modifications of fragmentation
functions based on radiative energy loss
\cite{Borghini:2005em,Armesto:2007dt,Armesto:2009fj,Renk:2009nz} have
generally predicted substantial reduction in the yield of high \pT,
or large-$z$ fragments and an enhancement at low \pT\ or low
$z$. The predicted reduction at large $z$ generically
results from the radiative energy loss of the leading partons in the
shower and the resulting redistribution of the jet energy to lower $z$
hadrons. Instead of a reduction, an enhanced yield of high $z$
fragments is seen in the data. However, the difference between
observed behaviour at large $z$ and expectations from theoretical
calculations may be at least partially attributed to the fact that the
fragmentation functions presented in this paper were evaluated with
respect to the energies of quenched jets. In contrast, theoretical
analyses of the fragmentation functions of quenched jets are typically
evaluated in terms of the initial, unquenched jet energies. 
However, some recent theoretical analyses \cite{Zapp:2012ak,Kharzeev:2012re} 
of jet fragmentation functions using quenched jet energies have 
shown that jet quenching calculations can reproduce the general
features observed in the results presented in this Letter. In addition
to direct modifications of the fragmentation function due to
quenching, the quenching may indirectly alter the fragmentation
function of inclusive jets by altering the relative fraction of quarks
and gluons.

The
simultaneous effects of quenching on the hadron constituents of jets
and the measured jet energies may explain a relative increase of
experimental fragmentation functions in central collisions at large
$z$ as suggested by the data. Jets that fragment to large-$z$ hadrons
may lose less energy than typical jets due to reduced formation or
colour-neutralization time \cite{KopNied1985}. Thus, the fragmentation
function  measured for inclusive jets may have a higher proportion of
jets with large-$z$ hadrons. The results in Ref.~\cite{Zapp:2012ak}
indicate such an effect that is qualitatively similar to the data.

\section{Conclusions}
This \papertype\ has presented measurements by ATLAS of charged-particle 
fragmentation functions in jets produced in $\sqrtsnn = 
2.76$~\TeV\ \PbPb\ collisions at the LHC. The measurements were
performed using a data set recorded in 2011 with
an integrated luminosity of $0.14~\invnb$. Jets were reconstructed
with the \antikt\ algorithm for distance parameters $R =0.2, 0.3,$
and 0.4, and the contributions of the underlying event to the jet
kinematics and the jet fragment distributions were
subtracted.  Jet fragments were measured within an angular range
$\Delta R = 0.4$ from the jet axes for all three jet sizes. 
Distributions of per-jet charged-particle transverse momentum,
\Dpt,  and longitudinal momentum fraction, \Dz, were presented for
seven bins in collision centrality for jet $\pT > 85, 92,$ and
100~\GeV, respectively, for \RTwo, \RThree, and \RFour\ jets. Ratios of
fragmentation functions in the different centrality bins to the
\mbox{60--80\%} bin were presented and used to evaluate the medium
modifications of jet fragmentation. Those ratios show an
enhancement in fragment yield in central collisions for $z
\lesssim 0.04$, a reduction in fragment yield for $0.04 \lesssim z
\lesssim 0.2$ and an enhancement in the fragment yield for $z > 0.4$. 
The modifications decrease monotonically with decreasing collision
centrality from \mbox{0--10\%} to \mbox{50--60\%}. A similar set of modifications is
observed in the \Dpt\ distributions over corresponding
\pT\ ranges.

\section*{Acknowledgements}

We thank CERN for the very successful operation of the LHC, as well as the
support staff from our institutions without whom ATLAS could not be
operated efficiently.

We acknowledge the support of ANPCyT, Argentina; YerPhI, Armenia; ARC,
Australia; BMWFW and FWF, Austria; ANAS, Azerbaijan; SSTC, Belarus; CNPq and FAPESP,
Brazil; NSERC, NRC and CFI, Canada; CERN; CONICYT, Chile; CAS, MOST and NSFC,
China; COLCIENCIAS, Colombia; MSMT CR, MPO CR and VSC CR, Czech Republic;
DNRF, DNSRC and Lundbeck Foundation, Denmark; EPLANET, ERC and NSRF, European Union;
IN2P3-CNRS, CEA-DSM/IRFU, France; GNSF, Georgia; BMBF, DFG, HGF, MPG and AvH
Foundation, Germany; GSRT and NSRF, Greece; ISF, MINERVA, GIF, I-CORE and Benoziyo Center,
Israel; INFN, Italy; MEXT and JSPS, Japan; CNRST, Morocco; FOM and NWO,
Netherlands; BRF and RCN, Norway; MNiSW and NCN, Poland; GRICES and FCT, Portugal; MNE/IFA, Romania; MES of Russia and ROSATOM, Russian Federation; JINR; MSTD,
Serbia; MSSR, Slovakia; ARRS and MIZ\v{S}, Slovenia; DST/NRF, South Africa;
MINECO, Spain; SRC and Wallenberg Foundation, Sweden; SER, SNSF and Cantons of
Bern and Geneva, Switzerland; NSC, Taiwan; TAEK, Turkey; STFC, the Royal
Society and Leverhulme Trust, United Kingdom; DOE and NSF, United States of
America.

The crucial computing support from all WLCG partners is acknowledged
gratefully, in particular from CERN and the ATLAS Tier-1 facilities at
TRIUMF (Canada), NDGF (Denmark, Norway, Sweden), CC-IN2P3 (France),
KIT/GridKA (Germany), INFN-CNAF (Italy), NL-T1 (Netherlands), PIC (Spain),
ASGC (Taiwan), RAL (UK) and BNL (USA) and in the Tier-2 facilities
worldwide.

\bibliography{JetFrag}
\bibliographystyle{elsarticle-num}


\onecolumn
\clearpage

\begin{flushleft}
{\Large The ATLAS Collaboration}

\bigskip

G.~Aad$^{\rm 84}$,
B.~Abbott$^{\rm 112}$,
J.~Abdallah$^{\rm 152}$,
S.~Abdel~Khalek$^{\rm 116}$,
O.~Abdinov$^{\rm 11}$,
R.~Aben$^{\rm 106}$,
B.~Abi$^{\rm 113}$,
M.~Abolins$^{\rm 89}$,
O.S.~AbouZeid$^{\rm 159}$,
H.~Abramowicz$^{\rm 154}$
% ...
\bigskip
\\
% ...
$^{11}$ Institute of Physics, Azerbaijan Academy of Sciences, Baku, Azerbaijan\\
% ...
$^{84}$ CPPM, Aix-Marseille Universit{\'e} and CNRS/IN2P3, Marseille, France\\
% ...
$^{89}$ Department of Physics and Astronomy, Michigan State University, East Lansing MI, United States of America\\
% ...
$^{106}$ Nikhef National Institute for Subatomic Physics and University of Amsterdam, Amsterdam, Netherlands\\
% ...
$^{112}$ Homer L. Dodge Department of Physics and Astronomy, University of Oklahoma, Norman OK, United States of America\\
$^{113}$ Department of Physics, Oklahoma State University, Stillwater OK, United States of America\\
% ...
$^{116}$ LAL, Universit{\'e} Paris-Sud and CNRS/IN2P3, Orsay, France\\
% ...
$^{152}$ Institute of Physics, Academia Sinica, Taipei, Taiwan\\
% ...
$^{154}$ Raymond and Beverly Sackler School of Physics and Astronomy, Tel Aviv University, Tel Aviv, Israel\\
% ...
$^{159}$ Department of Physics, University of Toronto, Toronto ON, Canada\\
% ...
\end{flushleft}


\end{document}